\documentstyle[prl,aps,preprint,tighten,floats,aps,epsf,psfig,axodraw]{revtex}

\def\be{\begin{equation}}
\def\ee{\end{equation}}
\def\bear{\begin{eqnarray}}
\def\eear{\end{eqnarray}}
\def\beqn{\begin{eqnarray}}
\def\eeqn{\end{eqnarray}}

\def\beq{\begin{equation} }
\def\eeq{\end{equation} }

\def\ben{\begin{eqnarray} }
\def\een{\end{eqnarray} }

\def\mod#1{{\rm (mod~2)} }

\def\Tr{{\rm Tr}}

\def\Tr{{\rm Tr}\,}

\def\ln{{\rm ln}\,}

\begin{document}
\draft
\preprint{\vbox{\baselineskip=12pt
\rightline{UPR-0814-T}
\vskip0.2truecm
\rightline{UM-TH-98/18}
\vskip0.2truecm
\rightline{hep-ph/9811355}}}

\title{Physics Implications of Flat Directions in Free Fermionic Superstring
Models II: Renormalization Group Analysis}
\author{
G. Cleaver${}^{\dagger}\footnote{Present address: Center for Theoretical
Physics, Texas A \& M University, College Station, Texas 77843-4242,
USA}$, 
M. Cveti\v c${}^{\dagger}$, 
J. R. Espinosa${}^{**}$, 
L. Everett${}^{\dagger}{}^{***}$, P. 
Langacker${}^{\dagger}$, and J. Wang${}^{\dagger}$}
\address{${}^{\dagger}$Department of Physics and Astronomy \\ 
          University of Pennsylvania, Philadelphia PA 19104-6396, USA \\
${}^{**}$CERN, TH Division\\
CH-1211 Geneva 23, Switzerland\\
${}^{***}$Randall Laboratory of Physics, University of Michigan\\
Ann Arbor, MI 48109, USA}
\maketitle
\begin{abstract}

We continue the investigation of the physics implications of a class of
flat directions for a prototype quasi-realistic free fermionic string
model (CHL5), building upon 
the results of the previous paper in which the complete
mass spectrum and effective trilinear couplings of the observable sector
were calculated to all orders in the superpotential.
We introduce soft supersymmetry breaking mass parameters into
the model, and investigate the gauge symmetry breaking patterns and the
renormalization group analysis for two representative flat directions,
which leave an additional $U(1)'$ as well as the SM gauge group unbroken
at the string scale.  We study symmetry breaking patterns 
that lead to a phenomenologically acceptable $Z-Z'$ hierarchy, $M_{Z^{'}}
\sim {\cal O}(1~{\rm TeV})$ and $ 10^{12}~{\rm GeV}$ for electroweak and
intermediate scale $U(1)^{'}$ symmetry breaking, respectively, and the
associated mass
spectra after electroweak symmetry breaking. The fermion mass spectrum
exhibits unrealistic features, including massless exotic fermions, but has
an interesting $d$-quark hierarchy 
and associated CKM matrix in one case. There are (some) non-canonical
effective $\mu$ terms, which lead to a non-minimal Higgs sector with more
than two Higgs doublets involved in the symmetry breaking, and a rich
structure of Higgs particles, charginos, and neutralinos, some of which,
however, are massless or ultralight. In the electroweak scale cases the
scale of supersymmetry breaking is set by the $Z^{'}$ mass, with the
sparticle masses in the several TeV range.  

\end{abstract}
\vskip2cm
\newpage
\section{Introduction}

In  recent work~\cite{cceel2,cceel3,cceelw1},
 techniques have been developed  which set the stage  for the 
``top-down'' analysis of   a class of
quasi-realistic string models.
Models in this class have $N=1$
supersymmetry, the SM gauge group as a part of the gauge structure, and
candidate fields for the three generations of quarks and leptons as well
as two electroweak Higgs doublets. Such quasi-realistic models have been
constructed in weakly coupled heterotic superstring theory in a variety of
constructions~\cite{orbifolds,calabiyau,NAHE,FNY1,AF1,chl}; in particular,
we consider a class of free fermionic models~\cite{NAHE,FNY1,AF1,chl}. 

In general, these quasi-realistic free fermionic models have an extended
gauge group (including a non-Abelian ``hidden" sector gauge group and a
number of additional $U(1)$'s), and a
large number of fields in addition to the MSSM fields, which include a
number of non-Abelian singlets, fields which transform under the hidden
sector gauge group, and SM exotics.  These models also have the
important property that the superpotential is calculable, in principle to
all orders in the nonrenormalizable terms.  The trilinear terms in the
superpotential have large Yukawa couplings of ${\cal O}(g)$, where $g$
is the gauge coupling at the string scale.  In contrast to
general field-theoretic models, additional string (worldsheet) symmetries
can forbid terms allowed by gauge invariance~\footnote{For
a review of the phenomenology of string models, see \cite{dienes} and
references therein.}.

These models generically possess  an ``anomalous''
$U(1)$ at the level of the effective theory. The
presence of the anomalous $U(1)$ leads to the generation of a
Fayet-Iliopoulos (FI) term at genus one; this term triggers scalar fields
to acquire string-scale VEV's along $D$ and $F$  flat directions, leading
to a ``restabilization" of the string vacuum.
Thus, as the necessary first step in the analysis, we developed
techniques~\cite{cceel2} to classify the
$D$ flat directions which can be proven to be $F$  flat to all orders of
a general perturbative heterotic superstring model with an
anomalous $U(1)$.  For the sake of simplicity, we chose to consider flat
directions formed of non-Abelian singlets only, and selected the singlet
fields with zero hypercharge to preserve the SM gauge group.  We applied
our method to a prototype string model, Model 5 of~\cite{chl} (CHL5),
in~\cite{cceel2}, and more recently to a number of free fermionic string
models in~\cite{cceel3}.

The next step in the analysis of this class of models is to analyze
the effective theory along such flat directions.  In general, the rank of
the gauge group is reduced, and effective couplings are
induced by the coupling of the fields in the flat direction to the rest of
the fields in the model.  The effective mass terms generated in this way
will give some of the fields superheavy masses, so that they decouple from
the theory at the string scale.  In addition to the trilinear
couplings of the original superpotential, effective trilinear terms
are generated from higher-dimensional terms for the remaining light
fields; such couplings can have important implications for the
phenomenology of the model~\footnote{The analysis of effective
nonrenormalizable terms  is deferred to further study. In this case
complications arise due to the fact that the fourth order 
(nonrenormalizable) terms present in the original superpotential could be 
competitive in strength \cite{CEW} with the trilinear ones, as  
well as generated in a number of other ways, such as via the decoupling of
heavy states\cite{ceew}, a nonminimal K\"{a}hler potential, and the
corrections to the K\"{a}hler potential due to the large VEV's. In the
following we will not include such effects and assume a minimal
K\"{a}hler potential.}.

In a recent work ~\cite{cceelw1} we studied in detail the physics 
implications of the effective theory after vacuum restabilization for 
a prototype model: Model 5 
of \cite{chl} (CHL5). The mass spectrum and the effective trilinear terms
(in the observable sector) were calculated exactly at the string tree level 
and to all orders in the vacuum expectation value of the fields in the flat 
directions for all of the flat directions classified in~\cite{cceel2}.
However, we presented the detailed analysis for
two representative flat directions, which encompass the general
features of this class of flat directions. These two flat
directions are part of a subset of 
flat directions of the model which break the maximal number of $U(1)$'s,
leaving an additional non-anomalous $U(1)'$ as well as the SM gauge group
unbroken at the string scale (as well as the hidden sector gauge groups).
Importantly, the string worldsheet symmetries forbid many of the 
gauge-allowed terms in the effective theory, which has a number of
implications for the phenomenology of the model.  

The calculation of the mass spectrum  revealed that along with the MSSM
fields, there remain a large number of massless exotic superfields at the
string scale, due to the absence of the corresponding effective mass terms
in the superpotential.  These fields include additional electroweak
doublets and electrically charged singlets, leading to a larger number of
fields in the observable sector in this model than in the MSSM.

While the scalar fields can acquire masses from the supersymmetry breaking, 
their fermionic superpartners can remain light compared with the
electroweak scale. We also found that many of the
massless exotic fields do not couple directly to the observable sector
fields at the order of the effective trilinear superpotential, such that
they do not participate in the radiative electroweak and $U(1)'$ gauge
symmetry breaking.  Therefore, there is no mechanism for these  exotic
fields to acquire masses. The
additional massless particle content has an impact on gauge coupling
unification, and renders the hidden sector gauge groups non-asymptotically
free (such that there is no possibility of dynamical supersymmetry
breaking in this sector due to strong coupling dynamics).

The presence of massless exotics  seems to be generic and thus it is a 
serious obstacle for deriving phenomenologically acceptable physics 
 from these models. Nevertheless,  they have enough realistic features that
it is worthwhile to study in detail the implications of the effective
trilinear terms of the superpotential. 
These  terms  have a  number of concrete implications for the observable
sector  physics.  In particular, the effective third order superpotential
does not 
have  a canonical $\mu$ term (involving the standard electroweak Higgs 
doublets) in either example, but instead non-canonical $\mu$ terms
involving some of the additional Higgs doublets. 
Such non-canonical terms suffice for the electroweak symmetry breaking.
However, in these examples,
there are not enough such terms to involve all of the Higgs doublets,
resulting in massless  charginos, neutralinos and an unwanted global
$U(1)$.  We found that when imposing the requirement that there are three
lepton doublets in the model, $R$- parity
is not conserved due to the presence of $L$- violating terms which leads
to possible lightest supersymmetric particle (LSP) decay in both of the
examples.  The second representative flat direction also has $B$-
violating terms (with implications for proton decay), as well as
couplings which yield textures in the quark and lepton sectors.

The purpose of this paper is to take the phenomenological analysis of
these models to the next stage: to determine the low energy 
implications of these string vacua for the gauge symmetry breaking
patterns and the analysis of the low energy mass spectra. Specifically, we 
investigate in detail the nature of the electroweak and $U(1)'$ gauge 
symmetry breaking patterns and the accompanying mass spectrum. Again, we 
choose to carry out this analysis for these two representative flat 
directions.

Such an analysis can only be performed after supersymmetry breaking has 
been incorporated. Since the origin of the supersymmetry breaking in string 
theory is not well understood, no quantitative, phenomenologically 
viable derivation of a supersmmetry breaking pattern is available. Thus we 
take a modest approach and parameterize the supersymmetry breaking by 
introducing soft supersymmetry breaking mass terms at the string scale. 
We can then proceed with the  study of  the renormalization group evolution 
of all the parameters in the observable sector and the study of their 
implications for the low-energy physics.


Although the  mass spectrum and the effective trilinear
superpotential for the representative flat directions of the CHL5 model
are not realistic due to the additional massless exotics
 and  the nature of the effective trilinear couplings (for details 
see ~\cite{cceelw1}), we choose to focus on the light fields which 
participate in the
gauge symmetry breaking, with the goal of obtaining scenarios
with a realistic $Z-Z'$ hierarchy.  As shown in \cite{cl}, in this class
of models the breaking scale of the $U(1)'$ can be either at the
electroweak (TeV) scale, or at an intermediate scale, depending on the
$U(1)'$ charges and trilinear couplings of the massless SM singlet fields.
In the first representative flat direction, the massless particle content
at the string scale does not allow for the intermediate scale $U(1)'$
breaking scenario, and hence the breaking is at the electroweak scale.
However, both scenarios are possible for the second representative flat
direction.

In this paper, we demonstrate these symmetry breaking scenarios explicitly 
for each of the two representative flat directions. Numerical results
are presented for specific (typical) choices of the soft supersymmetry 
breaking terms at the string scale, which in turn yield a realistic $Z-Z'$ 
hierarchy. We calculate the low energy spectrum explicitly with
the emphasis on the study of the Higgs mass spectrum and 
contrast its features with those of both the MSSM, and the ``bottom-up'' 
analysis of the string models with an additional $U(1)'$ studied
 in ~\cite{cdeel,lw}. 
(For the first flat direction the complete mass spectrum,
 including that of the supersymmetric partners, is presented.   
For the second flat direction we specifically address the texture in the 
(bottom) quark sector.) Due to the fact that the number of
 Higgs fields participating in the symmetry breaking pattern is larger
 than that assumed in \cite{cdeel},  the presence of some (but not all 
possible) non-canonical $\mu$ terms  
implies additional massless charginos, neutralinos, and Higgs bosons 
as well as new patterns in the  massive Higgs spectrum.

The paper is structured as follows.  In Section \ref{II}, we summarize the
features of the  CHL5 model  and define the two specific flat directions.
In Section \ref{III}, we  present the  
mass spectrum and the effective trilinear superpotential 
couplings   for the first  flat direction  of the CHL5 model.
 We demonstrate the possibility of realistic electroweak symmetry breaking
scenarios of the SM gauge group along with  the additional $U(1)'$. 
 In Section \ref{IV} we demonstrate the  $U(1)'$ 
  symmetry breaking scenarios at both the electroweak and intermediate
scales  for the second representative  flat direction. Finally, in Section
\ref{V} we present the summary and conclusions.

\section{Preliminaries}
\label{II}
The starting point  of the analysis is the effective theory along the two
representative flat directions of the CHL5 model~\cite{chl}. 
In previous work, we have presented techniques for classifying the flat
directions~\cite{cceel2} and demonstrated the calculation of the
effective couplings along these ``restabilized vacua'' of the
model~\cite{cceelw1}.  For the sake of completeness, we summarize the
method and the results here, and refer the reader
to \cite{cceel2,cceelw1} for the details.

\subsection{Method}
In the class of quasi-realistic string models considered, there is an
anomalous $U(1)$ generically present as part of the gauge structure, for
which the anomalies are cancelled by the four-dimensional version of the
Green Schwarz mechanism.  This standard anomaly cancellation mechanism
leads to the generation of a nonzero Fayet-Iliopoulos (FI) contribution
$\xi$ to the $D$ term of $U(1)_A$, with  
\begin{equation}
\xi = \frac{g^2_{\rm string}M_P^2}{192\pi^2}\Tr Q_A\, ,
\label{fid}
\end{equation}
in which $g_{\rm string}$ is related to
the  gauge coupling $g$ by the relation $g_{\rm
string}=g/\sqrt{2}$~\cite{Kap}
  ($g$ is  normalized according to the standard (GUT)  conventions,
 i.e., ${\rm Tr}T_aT_b=\delta_{ab}/2$ for the generators of the
fundamental representation of $SU(N)$) and $M_P= M_{Pl}/\sqrt{8\pi}$ is
the reduced Planck mass, with $M_{ Pl} \sim 1.2\times
10^{19}$ GeV.
This term would appear to break
supersymmetry in the original string vacuum. However, it triggers certain 
scalar fields to acquire VEV's of ${\cal O}(M_{String})$ along $D$ and
$F$ flat directions, leading to a supersymmetric ``restabilized" string
vacuum.  

Therefore, the classification of the flat directions is the
necessary first step in the analysis of the string model.
In \cite{cceel2}, we presented techniques to classify a subset of $D$ 
flat directions which can be proven to be $F$ flat to all orders in the
superpotential.  For the sake of simplicity and to preserve the SM gauge
group at the string scale, we chose to analyze flat directions formed from
non-Abelian singlet fields with zero hypercharge.  In general, the FI term
sets the scale of the VEV's in the flat direction, although in some cases
some of the VEV's are undetermined (but bounded from above).

The next stage of the analysis is to determine the effective theory along
such flat directions~\cite{cceelw1}.  In general, the rank of the gauge
group will be reduced (i.e., several $U(1)$'s will be broken).
Effective couplings can be generated from higher-dimensional operators in
the superpotential after replacing the fields in the flat direction by
their VEV's.  In particular, for a given flat direction $P$, effective
mass terms for the fields 
$\Psi_i$, $\Psi_j$ ($\Psi_{i,j}\notin  \{\Phi_k \in P\}$)
may be generated via 
\begin{equation}
\label{effmass}
W\sim\Psi_i\Psi_j\left(\Pi_{i\in P}\Phi_i\right)\ .
\end{equation}
In addition to these mass terms arising from $F$ terms, the fields in the
flat direction with VEV's set by the FI term will acquire masses
from $D$ terms via the superHiggs mechanism. The fields with such
effective mass terms will acquire string-scale
masses and decouple from the theory.  

In addition to the trilinear couplings of the original superpotential,
effective renormalizable interactions for the light fields may also be
generated
via
\begin{equation}
\label{efftril}
W\sim\Psi_i\Psi_j\Psi_k\left(\Pi_{i\in P} \Phi_i\right)\ .
\end{equation}
The existence of such terms was determined first
by identifying the gauge invariant effective bilinear and trilinear terms,
then subsequently verifying that  such  gauge invariant terms  survive the
string selection rules.

The coupling strengths  of the effective trilinear terms generated in
this way will generally be suppressed compared to the large Yukawa
couplings of the original superpotential, which have coupling strengths
${\cal O}(g)$ (with the typical value given by $\sqrt{2}g_{\rm
string}=g\sim 0.8$ for the models
considered, as is discussed in section III). 
In general, the coefficients of the superpotential
terms of order $K+3$ are given by
\begin{equation}
\label{alphadef}
\frac{\alpha_{K+3}}{M_{Pl}^K}=
g_{\rm string}
\left(\sqrt{\frac{8}{\pi}}\right)^{K}\frac{C_{K}I_{K}}{M^{K}_{Pl}}\ ,
\end{equation}
where $C_{K}$ is a coefficient of ${\cal O}(1)$ which includes
different renormalization factors in the operator product expansion (OPE)
of the string vertex operators (including the target space gauge group
Clebsch-Gordan coefficients), and $I_{K}$ is a world-sheet integral.
$I_{1,2}$ for certain typical couplings, have been computed numerically by several authors, with the typical 
result $I_1 \sim 70$, $I_2 \sim 400$ \cite{C}.  The coupling
strengths of the effective trilinear terms depend on these coefficients
and the values of the VEV's involved, which are set by the
(model-dependent) FI term.  In this way, these couplings can naturally
provide a hierarchy, with implications for generating fermion textures in
the quark and lepton sectors.

\subsection{Results: Model CHL5}
The model we have chosen as a prototype model to analyze is Model 5 of
\cite{chl}. Prior to vacuum restabilization, the model has the gauge
group
\begin{equation}
\{SU(3)_C\times SU(2)_L\}_{\rm obs}\times\{SU(4)_2\times SU(2)_2\}_{\rm hid}
\times U(1)_A\times U(1)^6, 
\end{equation}
and a particle content that includes the following  
chiral superfields in addition to the MSSM fields:
\begin{eqnarray}
&&6 (1,2,1,1) + (3,1,1,1) + (\bar{3},1,1,1) + \nonumber\\
&&4 (1,2,1,2) + 2 (1,1,4,1) + 10 (1,1,\bar{4},1) +\nonumber\\
&&8 (1,1,1,2) + 5 (1,1,4,2) + (1,1,\bar{4},2) +\nonumber\\
&& 8 (1,1,6,1) + 3 (1,1,1,3)+ 42 (1,1,1,1)\;\;,
\end{eqnarray}
where the representation under $(SU(3)_C,SU(2)_L,
SU(4)_2,SU(2)_2)$ is indicated.   We refer the reader
to~\cite{chl,cceel2,cceelw1} for the complete list of
fields with their $U(1)$ charges.

The SM hypercharge is determined as a linear combination of the six
non-anomalous $U(1)$'s, subject to the conditions that the MSSM fields
have the appropriate quantum numbers, and that the remaining fields can be
grouped into mirror pairs under $U(1)_Y$ (in the attempt to avoid the
presence of strictly massless colored or charged fermion fields in the theory).
In this model, these criteria lead to a unique definition of $U(1)_Y$
\cite{chl,cceelw1}, with Ka\v c-Moody level $k_Y=\frac{11}{3}$ (to be
compared with the MSSM value of $k_Y=\frac{5}{3}$). 

We presented a complete list of the $D$ flat
directions which can be proven to be $F$ flat to all orders in the
superpotential in~\cite{cceel2}, and
written more explicitly in Table II in ~\cite{cceelw1}. 
In~\cite{cceelw1}, we analyzed the $P_1P_2P_3$ subset of the flat 
directions, which includes the two representative directions
$P_1'P_2'P_3'$ and $P_2P_3|_F$.
The maximum number of $U(1)$'s is broken in these flat directions, which 
leave an additional $U(1)'$ as well as $U(1)_Y$ unbroken~\cite{cceel2}.
The unbroken $U(1)'$ has  
 $k_{Y'}=4167/250\simeq 16.67$.
The complete list of fields with their $U(1)_Y$ and $U(1)'$ charges is
presented in Tables Ia-Ic.

The VEV's of the fields in the most general 
$P_1P_2P_3$ flat direction are of the form
\be
\label{vevs}
\begin{array}{lcl}
|\varphi_{27}|^2=2x^2, &\;\;\;\;&
|\varphi_{28\,(29)}|^2=x^2-|\psi_1|^2,\nonumber\\
|\varphi_{30}|^2=|\psi_1|^2, &\;\;\;\; &
|\varphi_{4\,(5)}|^2=|\psi_2|^2,\nonumber\\
|\varphi_{2\,(3)}|^2=|\psi_1|^2-|\psi_2|^2, &\;\;\;\; &
|\varphi_{12\,(13)}|^2=|\psi_1|^2-|\psi_2|^2,\nonumber\\
|\varphi_{10\,(11)}|^2=|\psi_2|^2,&\;\;\;\;&
\end{array}
\ee
with 
\be
\label{xdef}
x=\frac{\sqrt{|\xi |}}{8}=0.013M_{\rm Pl},
\ee
and $|\psi_{1,2}|$ are free VEV's of the
moduli space, subject to the restrictions that
$x^2\ge|\psi_1|^2\ge|\psi_2|^2$. 

The first representative flat direction $P_1'P_2'P_3'$ has VEV's given
by the general case (\ref{vevs}), with no further
restrictions on $|\psi_{1,2}|$.  However, the second representative flat
direction $P_2P_3|_F$ corresponds to the case in which constraints must be
imposed on the free VEV's in such a way that the contributions from
different $F$ terms cancel and $|\varphi_{28\,(29)}|^2=0$; these
constraints are $x^2=|\psi_1|^2=2|\psi_2|^2$, and 
$\pi$  phase difference  between 
the combination of VEV's   $\varphi_{4\,(5)}\varphi_{10\,(11)}$ and 
$\varphi_{2\,(3)}\varphi_{12\,(13)}$~\cite{cceel2}.

For each of these flat directions, we computed the effective mass terms
and determined the mass eigenstates in~\cite{cceelw1}.  In addition to the
fields which become massive from these couplings, it can be shown that 
of the fields in each flat direction, five of the associated chiral
superfields  become  massive due to the superHiggs
mechanism. In the $P_1'P_2'P_3'$ flat direction,  two of the chiral
superfields remain massless (moduli), but they do not couple to the rest
of the fields at the level of the effective   trilinear terms. In the
$P_2P_3|_F$ flat direction,  the remaining complex  field gets a mass of
order [Yukawa] $\times$ [field VEV] due to cancellations of $F$ term
contributions and forms, along with its superpartner, a massive chiral
superfield. In this case, the fields with zero VEV's which couple
linearly in these terms also acquire masses of the same order.

The effective trilinear couplings along each of the representative flat
directions were also determined in~\cite{cceelw1}.  
In this model, the numerical  analysis (see also \cite{CEW}) of
the string amplitudes yields, along with the $x$ as the typical
VEV of the fields along the flat direction,  the effective  
Yukawa coupling  at the fourth  order $\sim 0.8$, while the fifth order
terms have  strengths $\sim 0.1$ (using the typical values of $I_1\sim70$ and
$I_2\sim400$ \cite{C}.)
Therefore, in this model the effective trilinear couplings arising from 
fourth order terms are competitive in strength to the elementary trilinear
terms, while the higher order contributions are  indeed suppressed
\cite{CEW}. 
Of course, the precise values for
each term will depend on the particular fields involved. In addition, 
the coupling strengths can depend on the undetermined VEV's
in the $P_1'P_2'P_3'$ flat direction.

\section{$P_1'P_2'P_3'$ Flat Direction}
\label{III}
\subsection{Effective Superpotential}

The $P_1'P_2'P_3'$ direction involves the set of fields
$\{\varphi_{2},\varphi_{5},\varphi_{10},\varphi_{13}, 
\varphi_{27},\varphi_{29},\varphi_{30}\}$.   The VEV's correspond to
the most general case given in (\ref{vevs}), such that they depend on two
free (but bounded) parameters.  

It is straightforward to determine the mass eigenstates, which were calculated
in~\cite{cceelw1}. In Table II we list the  surviving massless states.
 These states include both the usual  MSSM states and  
related exotic (non-chiral under $SU(2)_L$) states,
such as a fourth ($SU(2)_L$ singlet) down-type quark,
extra fields with the same quantum numbers as the lepton singlet superfields,
and extra Higgs doublets.
There are other massless states with exotic quantum numbers (including
fractional electric charge), and states which are non-Abelian representations
under both the hidden and observable sector gauge groups and thus directly mix
the two sectors.  
As previously discussed, there are  two additional massless states
(moduli) associated with fields which appear
in the flat direction but which do not have fixed VEV's (but do not couple to
other fields at the level of an effective trilinear term) and 
 are not listed in Table II.

The  effective trilinear couplings involving
the observable sector fields assume the form~\cite{cceelw1}:
\begin{eqnarray}
\label{superpot1}
W_{3}&=&gQ_cu^c_c\bar{h}_c+gQ_cd^c_bh_c+
\frac{\alpha_{4}^{(4)}}{M_{Pl}}\sqrt{1-\lambda_1^2}Q_c d_d^c h_a
+{g\over {\sqrt{2}}}e^c_ah_ah_c+
{g\over {\sqrt{2}}}e^c_fh_dh_c \nonumber\\&+&
\frac{\sqrt{2} \alpha^{(1)}_{5}x^2}{M_{Pl}^2}\lambda_2
e^c_h h_eh_a
+\frac{\sqrt{2} \alpha^{(2)}_{5}x^2}{M_{Pl}^2}
\sqrt{\lambda_1^2-\lambda_2^2} \
e^c_eh_eh_a 
+g\bar{h}_ch_b'\varphi_{20}',
\end{eqnarray}
in which 
\begin{equation}
\label{lambdadef}
\lambda_2\equiv\frac{|\psi_2|}{x} \le \lambda_1\equiv\frac{|\psi_1|}{x} \le 1,
\end{equation}
are free parameters and $\varphi_{20}'=\frac{1}{\sqrt{1+r^2}}
(\varphi_{20}-r\varphi_{22})$, with
$r \equiv [\alpha_4^{(1)} \lambda_2^2 + \alpha_4^{(2)}
(\lambda_1^2-\lambda_2^2)] x   
/(\sqrt{2}g M_{Pl})$.

The superpotential implies generic 
features independent of the details of the soft supersymmetry breaking, which
have been analysed in~\cite{cceelw1}:  
(i) With the identification of the fields  $\bar{h}_c$ and $h_c$ with 
the standard electroweak Higgs doublets, the Yukawa couplings indicate
$t-b$ and $\tau
-\mu$ Yukawa unification with equal  string scale Yukawa couplings $g$ and
$g/\sqrt{2}$, respectively; (ii) there is no  elementary or effective canonical
$\mu$-term; (iii) there  is a possibility of lepton- number violating
couplings and thus no stable LSP.  In particular, we identify the fields 
$\{h_e,h_a,h_d\}$ as the lepton doublets, and hence the couplings
$Q_cd_d^ch_a$ and $e^c_{e,h}h_ah_e$ violate lepton number. The fields $h_b'$ and
$\varphi^{'}_{20}$ can play the role of additional Higgs fields. 
$\bar{h}_{a}$ has the quantum numbers of a Higgs doublet, but does not
enter $W_{3}$, and hence there is no mechanism for it to develop a VEV.

\subsection{Symmetry Breaking Patterns}

To address the gauge symmetry breaking scenarios for this model, we
introduce soft supersymmetry breaking mass parameters 
, and
run the RGE's from the string scale to the electroweak scale.  
While the qualitative features of the analysis are independent of the
details of the soft breaking, we choose to illustrate the analysis with a
specific example with a realistic $Z-Z'$ hierarchy.

We wish to investigate the $U(1)'$ symmetry breaking
scenarios discussed in \cite{cl,cdeel,lw,cceel1}, which indicate that in
the class of string models considered, the $U(1)'$ symmetry breaking is
either at the electroweak (TeV) scale, or at an intermediate scale (if the
symmetry breaking
takes place along a $D$ flat direction).   
 An inspection of the massless spectrum in Table Ic indicates that the
singlet field $\varphi_{25}$ is required for a $D$ flat direction (and
hence the intermediate scale
$U(1)'$ symmetry breaking scenario); however, this field
acquires a string-scale mass for this direction, and decouples from the
theory.  We conclude that in this case, an intermediate scale breaking
scenario is not possible, and hence the breaking of the $U(1)'$ is
necessarily at the electroweak scale. 

Though hidden sector non-abelian fields are not directly or indirectly
coupled to the observable sector non-abelian fields at the
trilinear order in the superpotential, the $U(1)'$ could be radiatively
broken along with some of the hidden non-Abelian groups, and in such
cases the breaking of the hidden sector is connected with the $SU(2)
\times U(1)_{Y}$ symmetry breaking in the observable sector through
$U(1)^{'}$.

As discussed in \cite{cl,cdeel,lw}, several scenarios exist which can lead
to the possibility of a realistic $Z-Z'$ hierarchy. 
The scenario in which only the two MSSM Higgs fields $h_c$, $\bar{h}_c$
acquire VEV's breaks both $U(1)_Y$ and $U(1)'$, 
but leads to a light $Z'$ with $M_{Z^{'}} \sim {\cal O}(M_Z)$, which is
already excluded by experiments.   

To have a realistic $Z-Z'$ hierarchy, we require that a SM singlet field
that is charged under the $U(1)'$ acquires a VEV. However, since the 
canonical $\mu$ term that couples both $h_{c}$ and
$\bar{h}_{c}$ to a SM singlet is absent, and instead there is a
non-canonical $\mu$ term $\bar{h}_{c}h_{b}^{'}\varphi_{20}^{'}$, the
minimization of the potential requires that the
additional Higgs doublet $h_b'$ acquires a non-zero VEV. Therefore, we
consider the most general case in which $h_b'$ and $\varphi_{20}' \equiv
s$ acquire VEV's in addition to $\bar{h}_c$ and
$h_c$.  After adding the required soft supersymmetry breaking terms, the
potential is given by
\begin{eqnarray}
\label{potential}
V=V_F+V_D+V_{soft},
\end{eqnarray}
with
\begin{eqnarray}
\label{fterm1}
V_F&=&\Gamma_{s}^2|s|^2(|\bar{h}_c|^2+|h_b'|^2)+\Gamma_{s}^2|\bar{h}_c
\cdot
h_b'|^2,
\end{eqnarray}
\begin{eqnarray}
\label{dterm1}
V_D&=&\frac{G^2}{8}(|\bar{h}_c|^2-|h_c|^2-|h_b'|^2)^2+ 
\frac{g_2^2}{2}[|\bar{h}_c^{*}h_b'|^2+|\bar{h}_c^{*}h_{c}|^2+
|h_c^{*}h_b'|^2]\nonumber\\&-&\frac{g_2^2}{2}|h_b'|^2|h_{c}|^2 
+\frac{g_1'^2}{2}(Q_1|h_c|^2+Q_2|\bar{h}_c|^2+Q_3|h_b'|^2+Q_s|s|^2)^2,
\end{eqnarray}
\begin{eqnarray}
\label{vsoft}
V_{soft}&=&m^2_{\bar{h}_c}|\bar{h}_c|^2+m^2_{h_c}|h_{c}|^2+m^2_{h_b'}|h_b'|^2
+m^2_{s}|s|^2\nonumber\\&-&(A\Gamma_{s}\bar{h}_{c}\cdot h_b' s +h.c.),
\end{eqnarray}
in which $\Gamma_{s}$ is the coefficient for the coupling $\bar{h}_{c} h_{b}' \varphi_{20}'$, $G^2=g_Y^2+g_2^2$ (with $g_Y^2=\frac{3}{11}g_1^2$), and 
\begin{eqnarray}
h_c=\left(\begin{array}{c c} h_{c}^0\\h_c^-\end{array}\right),\;
\bar{h}_{c}=\left(\begin{array}{c c}
\bar{h}_c^+\\\bar{h}_c^0\end{array}\right), \;
h_b'=\left(\begin{array}{c c} h_b'^0\\h_b'^-\end{array}\right).
\end{eqnarray}
The $U(1)'$ charges of $\{h_{c},\bar{h}_c,h_b',s\}$ are denoted by
$Q_1$, $Q_2$, $Q_3$, and $Q_s$, respectively.
We can take $A\Gamma_s$ real and positive without loss of generality by an
appropriate choice of the global phases of the fields.  
By a suitable gauge rotation we 
also take $\langle \bar{h}_{c}^{0} \rangle$ and $\langle s \rangle$ 
real and positive, which implies that $\langle h_b^{'0} \rangle$ is real
and positive at the minimum. However, the phase of the $h_c$ 
field is not determined, due to the absence of an effective  $\mu$   
term involving $h_c$ in (\ref{potential}).  This additional global $U(1)$
symmetry leads to the presence of a Goldstone boson in the
massless spectrum, as discussed below.  For
notational
simplicity, we define $\sqrt{2} \langle h^0_c \rangle \equiv v_1$,
$\sqrt{2}\langle
\bar{h}^0_c \rangle \equiv v_2$, $\sqrt{2}\langle h_b^{'0} \rangle \equiv 
v_3$, and $\sqrt{2} \langle s \rangle \equiv s$.

The $Z-Z'$ mass matrix is given by
\begin{eqnarray}
\label{Zmatrix}
(M^{2})_{Z-Z'}=\left (\begin{array}{c c}
M_{Z}^{2}&\Delta^{2}\\\Delta^{2}&M_{Z'}^{2}\end{array}\right),
\end{eqnarray}
where
\begin{eqnarray}
M_{Z}^{2}&=&\frac{1}{4}G^2(v_{1}^2+v_{2}^2+v_3^2),\\
M_{Z'}^{2}&=&{g'}_{1}^{2}(v_{1}^{2}Q_{1}^{2}+v_{2}^{2}Q_{2}^{2}+v_3^2 Q_{3}^{2}+s^{2}Q_{S}^{2}),\\
\label{mix}
\Delta^{2}&=&\frac{1}{2}g_{1}'\,G(v_{1}^2Q_{1}+v_{3}^2Q_{3}-v_{2}^2Q_{2});
\end{eqnarray} 
with mass eigenvalues
\begin{eqnarray}
M^{2}_{Z_{1},Z_{2}}&=&\frac{1}{2}\left[M^{2}_{Z}+M^{2}_{Z'}\mp
   \sqrt{(M^{2}_{Z}-M^{2}_{Z'})^{2}+4\Delta^4}\right].
\end{eqnarray}  
The $Z-Z'$ mixing angle $\alpha_{Z-Z'}$ is given by
\begin{eqnarray}
\alpha_{Z-Z'}=\frac{1}{2}\arctan\left(\frac{2\Delta^2}{M^{2}_{Z'}-M^{2}_{Z}}
\right),
\end{eqnarray} 
which is constrained to be less than a few times $10^{-3}$.

The only possibility~\cite{cceelw1} for a realistic
hierarchy is for the symmetry breaking to be characterized by a large
(${\cal O}({\rm TeV})$) value for the SM singlet VEV $s$, with the
$SU(2)_L \times U(1)_Y$ breaking at a lower scale due to accidental
cancellations.

We now proceed with the analysis of the renormalization group 
equations.\\

{ (i) Running of the Gauge Couplings:} \\




As discussed in \cite{cceelw1}, we determine the gauge coupling constant 
$g=0.80$ at the string scale by assuming $\alpha_{s}=0.12$ (experimental
value) at the electroweak scale, and evolving $g_{3}$ to the string scale.
We find that $g=0.80$ is slightly higher than that of the MSSM, due to the
presence of one additional vectorlike exotic quark pair. The electroweak
scale values of the other gauge couplings are determined by their 
(1-loop) RGE's, taking $g=0.8$ at the string scale as an input. The
running of the
gauge couplings is presented in Figure 1, and the $\beta$ functions are
listed in Table III, including the Ka\v c-Moody levels for the $U(1)$
gauge factors ($k_{Y}=11/3$, $k^{'}=16.67$) and $k=2$ for the
hidden sector non-Abelian groups. 

The low energy values of the gauge couplings are not correct due to the
exotic matter and non-standard $k_{Y}$.  
Surprisingly, $\sin^{2}{\theta_{W}} \sim 0.16$ is not too different from the
experimental  value $0.23$, and $g_{2}=0.48$ is to be compared to the
experimental value $0.65$. As a result of a large
number of massless non-Abelian fields in the hidden sector, the hidden
sector gauge couplings are not asymptotically free. Therefore,
dynamical supersymmetry breaking due to strong coupling dynamics 
in the hidden sector is not possible in this model. \\   

\begin{figure}
\vskip -0.3truein
\centerline{
\hbox{
\epsfxsize=3.5truein
\epsfbox[70 32 545 740]{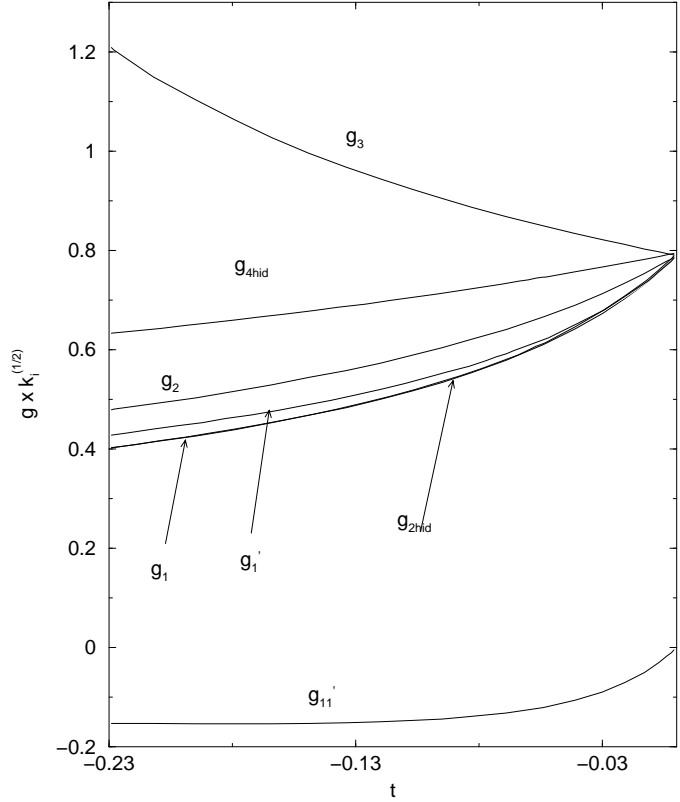}
\hskip 0.25truein
}
}
\caption{
Variation of the gauge couplings $\times \sqrt{k}$ with the scale for the
$P_1'P_2'P_3'$ flat direction, with $t=(1/16\pi^2)\ln(\mu/M_{String})$,
$M_{String}=5 \times 10^{17}$ GeV, and $g(M_{String})=0.80$.  The
couplings include the factor $\sqrt{k}$, where k corresponds to the
associated Ka\v c-Moody level (see the caption of Table III for the values
of $k$).}
\end{figure}

\begin{figure}
\centerline{
\hbox{
\epsfxsize = 3.5truein
\epsfbox[70 32 545 740]{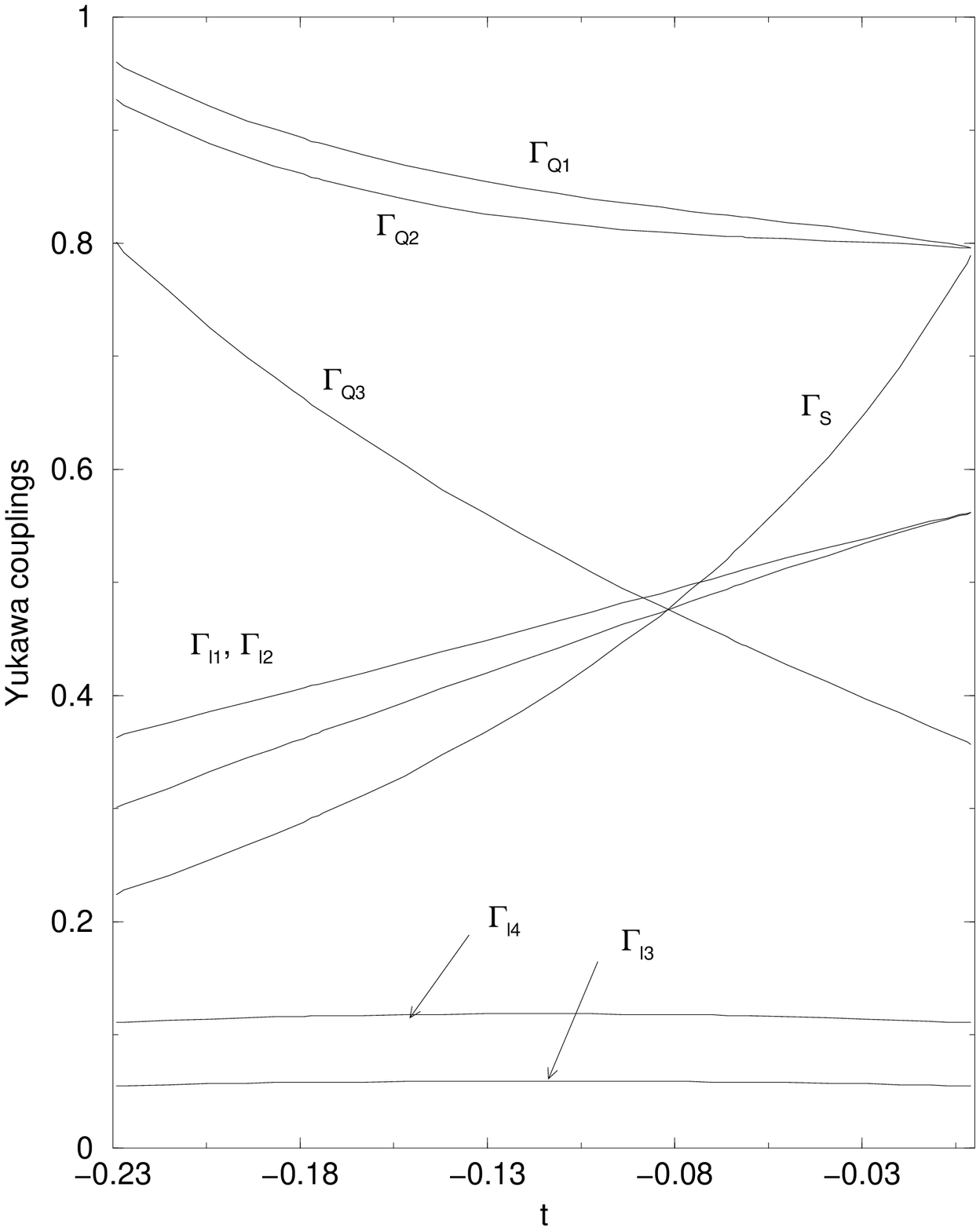}
}
}
\caption{
Running of the Yukawa couplings for the $P_{1}^{'}P_2^{'}P_3^{'}$ direction.
 The two free parameters $\lambda_{1}$ and $\lambda_{2}$
are chosen to be $0.9$ and $0.4$, respectively. 
}

\end{figure}

{ (ii) Running of the Yukawa Couplings:} \\

The values of the Yukawa couplings at the string scale after vacuum
restabilization are indicated in (\ref{superpot1}).
There are two free parameters $\lambda_{1}$ and
$\lambda_{2}$ satisfying the constraints (\ref{lambdadef}). 
To minimize the effects of the lepton number violating effective Yukawa
coupling $Q_c d_d^c h_a$, we choose $\lambda_1 =0.9$ at the string
scale. The dependence of $\lambda_2$ is all from higher order effective
terms, which are numerically supressed and less important. For the sake of
definiteness, we take   
$\lambda_{2}=0.4$. As discussed in section II, we choose the typical values 
for $I_{1,2}$ such that 
$\frac{\alpha_{4} x}{M_{Pl}} \sim 0.8$, $\frac{\alpha_{5}
x^2}{ M_{Pl}^{2}}\sim 0.1$, and an estimation for $I_{3}$
such that$\frac{\alpha_{6} x^3}{M_{Pl}^3} \sim 0.01$. 
With the choice of $g=0.8$ at the string  scale, the initial values of
Yukawa couplings are listed in Table IV. In
Fig. 2 we present the evolution of the Yukawa couplings with the scale.
We denote the Yukawa couplings of the quark doublets by $\Gamma_{Qi}$, the
couplings of the lepton doublets by $\Gamma_{li}$, the coupling of the two
Higgs doublets and the singlet by $\Gamma_s$ 
(the numbering follows the order of the terms in (\ref{superpot1})).
Additional Yukawa couplings of non-abelian singlet fields to hidden
sector   fields, not displayed in (9), are listed in eqn. (20) of
\cite{cceelw1} .
Their effects are included in the calculation of the running Yukawas and
  soft parameters.

The low energy values of the Yukawa couplings can be read off from Table
IV. The values of $\Gamma_{Q1}\sim 0.96$ and $\Gamma_{Q2}\sim 0.93$
indicate that the
$t-b$ degeneracy is mainly broken by $\tan{\beta}$ at the electroweak
scale.  The result $\Gamma_{l1} \sim 0.30$, $\Gamma_{l2} \sim 0.36$
indicates that the $\tau-\mu$ degeneracy is only slightly 
broken at low energy, and  $m_{b}/m_{\tau} \sim 2.6$. 
The value of the Yukawa coupling $\bar{h}_{c} h_{b}^{'}
\varphi_{20}^{'}$, which  plays a significant role in the gauge symmetry
breaking (in this scenario for which all three fields have non-zero
VEV's), is given by $\Gamma_{s}(M_Z) \sim0.22$.
\\

{ (iii) Running of the Soft Mass Parameters:}\\

We choose to normalize the soft breaking scale by ensuring
the correct value of $M_Z$ (before mixing with $Z'$), rather than using $v$ or
$M_W$; since in this modle the gauge couplings $g_2$ and $g_1$ at $M_{Z} $ 
have values different from their experimentally observed ones, it implies that
$v$ and $M_W$ will also differ from their experimentally measured values. 
With the (incorrect) values
$g_2(M_{Z})\sim0.48$, $g_1(M_Z)\sim 0.41$ for this model,
$v\sim 348$ GeV  and $M_W \sim 82.8$ GeV 
(to be compared with the experimental values 246 GeV and 80.3 GeV, 
respectively).

We find that with universal boundary  conditions for the soft supersymmetry
breaking mass terms at the string scale, the realistic scenario described
above cannot be achieved. In
particular, the mass-square of the appropriate singlet field $\varphi_{20}'$
does not run to negative values, so $\varphi_{20'}$ does not acquire a
VEV. However, for mild tuning of the boundary conditions
it is possible to obtain scenarios
in which the $Z'$ mass is large enough and the mixing angle sufficiently
suppressed to satisfy phenomenological bounds. 

We present the initial conditions and low energy values for the soft breaking
parameters for an example of
such a scenario in Table IV, with $M_{Z'}=735$ GeV, and $\theta_{Z-Z'}=0.005$. 
The running scalar mass-squares corresponding to these initial conditions
and to the gauge and Yukawa couplings in Figure 1 and Figure 2 are
displayed in Figure 3. 

\begin{figure}
\centerline{
\hbox{
\epsfxsize=3.5truein
\epsfbox[70 32 545 740]{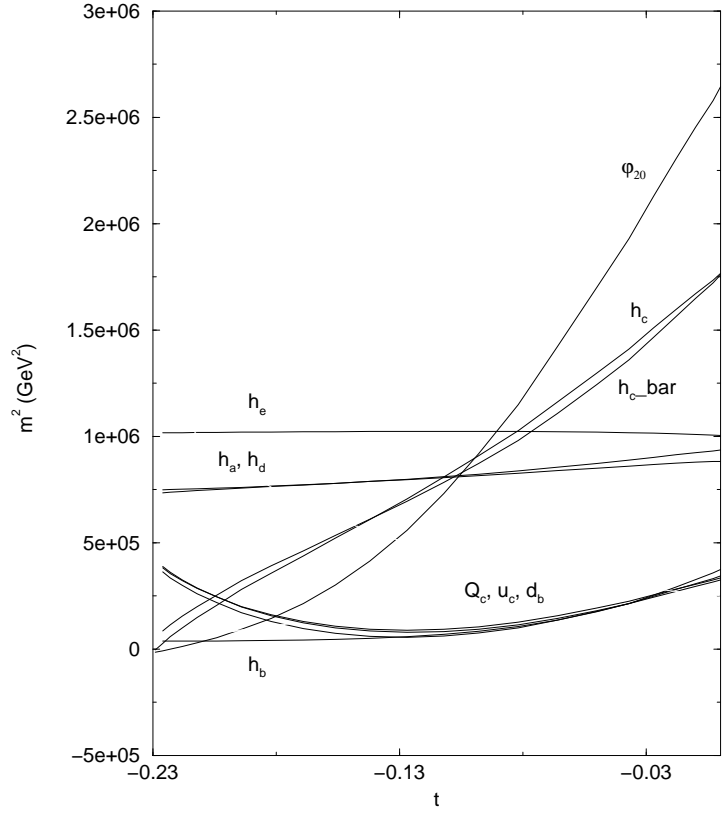}
}
}
\caption{
Running of the soft mass-squared parameters for
$P_{1}^{'}P_{2}^{'}P_{3}^{'}$ direction. The non-universal initial
conditions are chosen to yield a realistic example for low energy symmetry
breaking. 
}
\end{figure}

We see that the mass-squares of $h_c$ and  
$\varphi^{'}_{20}$ are driven negative at low energy, while the
$\bar{h}_c$ and $h_{b'}$ mass-squares
remain positive.  Minimization of the potential requires all four of these
fields to acquire VEV's, which take the values 
$\langle h_c^0 \rangle = 90$ GeV,
$\langle \bar{h}_c^0 \rangle = 163$ GeV,
$\langle h_{b}^{'0} \rangle = 161$ GeV, and
$\langle \varphi_{20}' \rangle = 3560$ GeV. 


The small values of the doublet VEV's compared to $\langle \varphi_{20}'
\rangle$ involve a degree of fine-tuning. From Table IV it is apparent that
the typical scale of all soft parameters, and therefore of the VEV's and
$Z$ and $Z^{'}$ masses is several TeV. The much smaller values of
$\langle \bar{h}_c^0 \rangle$ and $\langle h_b^{'0} \rangle$ come about
because the point $\langle \bar{h}_c^0 \rangle= \langle h_{b}^{'0} \rangle
=0$ of the potential is a saddle point, with large positive curvature in
one direction, and a small negative curvature (caused by a near 
cancellation between the large positive $m_{\bar{h}_{c}}^{2}$ and
$m_{h_b'}^{2}$ terms with the slightly larger negative $A$ term), in
the other direction. The small $\langle h_c^0 \rangle$ is due to the small
negative $m_{h_{c}}^{2}$ and the absence of a trilinear term involving
$h_{c}$.   

\subsection{ \bf{Mass Spectrum} }

We now address the mass spectrum of the model associated with this
particular low-energy solution.  
The large singlet VEV 
scenario was explored for models with two 
Higgs doublets and a singlet connected by a canonical $\mu$ term in
\cite{cdeel}. In the present model, there are three Higgs doublets and one
singlet involved in the symmetry breaking, and one of the doublets ($h_c$)
does not have trilinear couplings to the singlet, so that it enters the
potential only through $D$ terms.  We find that while 
the pattern of the masses obtained generally follows the pattern obtained
in \cite{cdeel}, there are additional features in the mass spectrum of
the charginos, neutralinos, and Higgs scalars. \\


(1) Fermion Masses: \\ 

With the identification of $Q_c$ as the quark doublet
of the third family, $m_t=156$ GeV, and $m_{b}=83$ GeV,
where $m_b$ is evaluated at $M_Z$. The low value for $m_t$, despite the
high value of $v = 348$ GeV is because of the low value of
$\sqrt{2}\langle \bar{h}_c^0 \rangle /v \sim 0.66$. Clearly, $\langle
h_{c}^0 \rangle$ is much too large for this example~\footnote{Smaller and
more realistic values for $m_{b}$ and $m_{\tau}$ could have been obtained
by further adjustment of parameters to yield a smaller $\langle h_{c}^{0}
\rangle$. This would violate our strategy of presenting a typical model
with a realistic $Z-Z^{'}$ hierarchy without further adjustment. Smaller
$\langle h_{c}^{0} \rangle$ would also have resulted in still smaller
masses for the lightest charged and scalar Higgs particles.} 
as is reflected in the
unacceptably large value for $m_{b}$. That is, the $t-b$ unification is not 
acceptable in the example because it would require a large ratio of
 $\langle \bar{h}_c^0 \rangle / \langle h_{c}^0 \rangle $. If we identify
 $h_d$, $h_a$ as the lepton doublets of the third and second families, 
 $m_{\tau}=32$ GeV, and $m_{\mu}=27$ GeV, where the difference
 is due to the $e^c_{h,e} h_e h_a$ terms in $W_3$.
 The ratio $m_b/m_\tau$ is larger than in the usual $b-\tau$ unification
 because of the ratio $1:1/\sqrt{2}$ of the Yukawa couplings at the string
scale, and is probably inconsistent with experiment \cite{mbmtau}. Of
 course, the high value for $m_\mu$ is unphysical. 

 There is no mechanism to generate
 significant $u$, $d$, $c$, $s$, and $e^-$ masses for this direction. \\

(2) Squarks/Sleptons: \\ 

The squark and slepton masses take the values 
$m_{\tilde{t}\,L}=2540$ GeV, 
$m_{\tilde{t}\,R}=2900$ 
GeV; $m_{\tilde{b}\,L}=2600$ GeV, $m_{\tilde{b}\,R}=2780$ GeV;
$m_{\tilde{\tau}\,L}=2760$ GeV, $m_{\tilde{\tau}\,R}=3650$ GeV; 
$m_{\tilde{\mu}\,L}=2790$ GeV, $m_{\tilde{\mu}\,R}=3670$ GeV.
The large values are needed to ensure a large $M_{Z'}$ in this
model. The numerical values actually refer to the mass eigenstates,
which are mixtures of the $L$ and $R$ states. However, the $L-R$
mixing terms are small compared to the diagonal terms for this
example, so the mixing effects are small.
The other squark and slepton masses depend
on initial values for the soft supersymmetry breaking mass parameters that
approximately decouple from the symmetry breaking pattern, and are not
presented. \\

(3) Charginos: \\

The positively charged gauginos and higgsinos are 
$\{\tilde{W}^{+}, \tilde{\bar{h}}_c, \tilde{\bar{h}}_a \}$, 
and the negatively charged  gauginos and higgsinos are $\{\tilde{W}^{-},
\tilde{h}_c,\tilde{h}_b' \}$.  The mass matrix is given by 
\begin{eqnarray}
\label{charginos}
M_{\tilde{\chi}^{\pm}}=\left(\begin{array}{c c c}
M_2 & {\displaystyle\frac{1}{\sqrt{2}}}g_{2}v_1 &
{\displaystyle\frac{1}{\sqrt{2}}}g_{2}v_3
\vspace{0.1cm}\\
{\displaystyle\frac{1}{\sqrt{2}}} g_{2}v_1 & 0 & \Gamma_s
{\displaystyle\frac{s}{\sqrt{2}}} \vspace{0.1cm}\\
0 & 0& 0 
\end{array}\right).
\end{eqnarray}
There is one massless chargino, and the
other two are massive, with masses 
$m_{\tilde{\chi}_{1}^{\pm}}= 591$ GeV, and 
$m_{\tilde{\chi}_{2}^{\pm}}= 826$ GeV. 

The massless state involves $\bar{h}_a$ (and a linear combination
of the negative states), and is due to the absence of
$\bar{h}_a$ couplings in the superpotential. In particular, there
is no non-canonical $\mu$ term $\bar{h}_a h_c \varphi$ in this flat
direction; the
only gauge-allowed
term of this type ($\bar{h}_a h_c \varphi_{25}$) was in fact in the
original trilinear
superpotential, but $\varphi_{25}$ has acquired a string-scale mass and
decoupled from the low-energy theory.  Therefore, there is no mechanism
for $\bar{h}_a$ to acquire a VEV. \\

(4) Neutralinos: \\

The neutralino sector consists of $\{\tilde{B}',\tilde{B},
\tilde{W_3},\tilde{\bar{h}}_{c}^0,\tilde{h}_{c}^0,
\tilde{h}_{b}^{'0},
\tilde{\varphi}^{'}_{20},\tilde{\bar{h}}_{a}^0\}$.
In this basis (neglecting $\tilde{\bar{h}}_{a}^0$, which has no
couplings), the neutralino mass matrix is given by
\begin{eqnarray}
\label{neutralinos}
M_{\tilde{\chi}^0}=\left(\begin{array}{c c c c c c c }
M'_1 & 0 & 0 &
 g'_1Q_2v_2 &
 g'_1Q_1v_1 &
 g'_1Q_3v_3 &
 g'_1Q_Ss \vspace{0.1cm}\\
0 & M_1 & 0 &
{\displaystyle\frac{1}{2}}g_Yv_2 &
-{\displaystyle\frac{1}{2}}g_Yv_1 & 
-{\displaystyle\frac{1}{2}}g_Yv_3 & 
0\vspace{0.1cm}\\
0 & 0 & M_2 &
-{\displaystyle\frac{1}{2}}g_2v_2 &
{\displaystyle\frac{1}{2}}g_2v_1 &
{\displaystyle\frac{1}{2}}g_2v_3 &
 0\vspace{0.1cm}\\
g'_1Q_2v_2 &
{\displaystyle\frac{1}{2}}g_Yv_2 &
-{\displaystyle\frac{1}{2}}g_2v_2 & 0 &0 
& -{\displaystyle\frac{1}{\sqrt{2}}}\Gamma_s s
& -{\displaystyle\frac{1}{\sqrt{2}}}\Gamma_s v_3
\vspace{0.1cm}\\
 g'_1Q_1v_1 &
-{\displaystyle\frac{1}{2}}g_Yv_1 &
{\displaystyle\frac{1}{2}}g_2v_1 &
0&0&0&0 \vspace{0.1cm}\\
 g'_1Q_3v_3 &
-{\displaystyle\frac{1}{2}}g_Yv_3 &
{\displaystyle\frac{1}{2}}g_2v_3 &
-{\displaystyle\frac{1}{\sqrt{2}}}\Gamma_s s& 0 &0&
 -{\displaystyle\frac{1}{\sqrt{2}}}\Gamma_s v_2\vspace{0.1cm}\\
 g'_1Q_S s & 0 & 0 &
-{\displaystyle\frac{1}{\sqrt{2}}}\Gamma_s v_3 & 0&
-{\displaystyle\frac{1}{\sqrt{2}}}\Gamma_s v_2 & 0
\end{array}\right).
\end{eqnarray}
The mass eigenvalues are:
$m_{\tilde{\chi}_{1}}^0= 963$ GeV, 
$m_{\tilde{\chi}_{2}}^0= 825$ GeV, 
$m_{\tilde{\chi}_{3}}^0= 801$ GeV, 
$m_{\tilde{\chi}_{4}}^0= 592$ GeV, 
$m_{\tilde{\chi}_{5}}^0= 562$ GeV, 
$m_{\tilde{\chi}_{6}}^0= 440$ GeV, 
$m_{\tilde{\chi}_{7}}^0= 2$ GeV,
 and $m_{\tilde{\chi}_{8}}^0= 0$.

$\tilde{\chi}_{8}^0$ corresponds to $\tilde{\bar{h}}_{a}^0$, which
(as previously mentioned) does not enter the superpotential.
The hierarchy of the non-zero masses 
can be understood in the large singlet VEV scenario $\Gamma_s^2
\langle \varphi^{'}_{20} \rangle^2 \gg M_i^2, M_Z^2$ (in which $M_i$ denotes
the gaugino masses), 
which  yields  the pattern:\\
$\tilde{\chi}_{1,2}=(\tilde{\bar{h}}_{c}^0 \pm \tilde{h}_{b}^{'0})/\sqrt{2}$, 
with masses $\sim \Gamma_s \langle \varphi^{'}_{20} \rangle$;
$\tilde{\chi}_{3,4}=(\tilde{B}' \pm \tilde{\varphi}^{'}_{20})/\sqrt{2}$, 
with masses $\sim M_{Z}'$;\\
$\tilde{\chi}_{5,6}=\tilde{B}',\tilde{W}^{0}$, with masses $\sim |M_1|$,
$|M_2|$;\\
$\tilde{\chi}_{7}=\tilde{h}_{c}^0$, with mass $\sim 0$. \\

(5) Exotics: \\

In addition to the massless quarks, leptons, chargino, and neutralino
discussed above, there are a number of exotic states, including
the $SU(2)_L$ singlet down-type quark, four $SU(2)_L$ singlets with
unit charge (the $e$ and extra $e^c$ states), and a number of
SM singlet ($\varphi$) states. There are additional exotics
associated with the hidden sector. The scalar components of these
exotics are expected to acquire TeV-scale masses by soft supersymmetry
breaking. However, there is no mechanism for this direction to give the
fermions a significant mass. In particular, fermion masses associated with
higher-dimensional operators would be suppressed by powers of the ratio of
the TeV scale to the
string scale, and are therefore negligible. (Such operators could
be a viable mechanism for other flat directions that allow
an intermediate scale, however.)
Another possible mechanism would be to invoke a non-minimal
K\"{a}hler potential. However, that is beyond the scope of the present
analysis. \\

(6) Higgs Sector: \\

The non-minimal Higgs sector of three complex doublets and one
complex singlet required for this scenario leads to additional Higgs
bosons compared to the MSSM.  In this scenario, four of the
fourteen degrees of freedom are  eaten to become the longitudinal
components of the $W^{\pm}$, $Z$, and $Z'$; in addition, there is a
global $U(1)$ symmetry present in (\ref{potential}) associated with the 
phase of $h_{c}$ which is broken, leading to
a massless Goldstone boson in the spectrum. It would acquire a small mass 
at loop level due to couplings in the full theory which do not respect 
the global $U(1)$.   

The spectrum of the physical Higgs bosons after symmetry breaking
consists of two pairs of charged Higgs bosons $H_{1,2}^{\pm}$, four
neutral CP even
Higgs scalars $(h^0_i\, ,i=1,2,3,4)$, and one CP odd Higgs $A^0$.

In the basis
$\{\bar{h}^{0\,i}_c
\equiv \sqrt{2}{\rm Im} \bar{h}_c^0, 
h_c^{0\,i}, h_b^{'0\,i}, s^{i}\}$), the CP odd (tree-level) mass matrix 
is given by
\begin{eqnarray}
M^2_{A^0}=\frac{A \Gamma_s}{\sqrt{2}}\left( \begin{array}{c c c c}
{\displaystyle\frac{sv_3}{v_2}} & 0&  s& v_3
\vspace{0.1cm}\\
0&0&0&0\\
s&0 & {\displaystyle\frac{sv_2}{v_3}}&
v_2
\vspace{0.1cm}\\
v_3 & 0&
v_2 & {\displaystyle\frac{v_2 v_3}{s}}. \end{array}\right)
\end{eqnarray}
There are one massive and three massless eigenstates.  Two of the
massless eigenstates are the Goldstone bosons which are absorbed to become
the longitudinal components of the $Z$ and the $Z'$.  The third massless
state is the Goldstone boson corresponding to the breakdown of
the global $U(1)$ symmetry present in (\ref{potential}), due to
the absence of trilinear couplings involving $h_c$.  The physical CP
odd Higgs has
\begin{eqnarray}
m^2_{A^0}=\frac{A\Gamma_s}{\sqrt{2}}(\frac{s v_2}{v_3}+  \frac{s v_3}{v_2}
+\frac{v_2 v_3}{s}),
\end{eqnarray}
which takes the value $m_{A^0}=1650$ GeV in this particular case.

The mass matrix of the charged Higgses in the basis $\{h_{c}^{-}$,
$\bar{h}_{c}^{+*}$, $h_{b}^{'-} \}$ takes the form
\begin{eqnarray}
M^2_{H^{\pm}} = \left( \begin{array}{c c c}
\frac{g_{2}^2}{4}(v_2^{2}-v_3^{2}) &
\frac{g_2^{2}}{4}v_1 v_2 &
\frac{g_2^{2}}{4}v_1 v_3 \\
\frac{g^2_{2}}{4}v_1 v_2 & A \Gamma_s \frac{ s
v_3}{\sqrt{2}v_2}+\frac{g^2_{2}}{4}(v^2_{1}+v^2_{3}) 
-\Gamma_{s}^{2}\frac{v^2_{3}}{2} &
\frac{g^2_{2}}{4}v_2 v_3 -
\Gamma_{s}^2 \frac{v_2 v_3}{2} + A \Gamma_s
\frac{s}{\sqrt{2}} \\
\frac{g_2^{2}}{4}v_1 v_3 &
\frac{g_2^{2}}{4}v_3 v_2
-\Gamma_s^{2}
\frac{v_2 v_3}{2} + A\Gamma_s
\frac{s}{\sqrt{2}} & A \Gamma_s \frac{sv_2}{\sqrt{2}v_3}+\frac{g_2^{2}}{4}
(v_2^{2}-v_1^{2})-\Gamma_s^{2} \frac{v_2^{2}}{2}
\end{array} \right)
\end{eqnarray}

There is one massless state, which is the Goldstone boson absorbed by
$W^{\pm}$ after the $SU(2)$ symmetry is spontaneously broken; the two
physical charged Higgses are $m_{H^{\pm}_1}=10$ GeV; $m_{H^{\pm}_2}=1650$
GeV. The extra light mass of $H^{\pm}_1$ is due to an accidental
cancellation between $v_2$ and $v_3$ for the specific example considered;
in general, it is $\sim {\cal O}(M_{Z})$.

The masses for the four neutral scalars can be obtained by diagonalizing
the mass matrix (in the basis $\{\bar{h}^{0\,r}_c,h^{0\,r}_c,
h_{b}^{'0\,r}, s^{r}\}$)
\begin{eqnarray}
M^2_{h^0}=\left(\begin{array}{c c c c}
\kappa_2^2v_2^2+A\Gamma_s
s{\displaystyle\frac{v_3}{\sqrt{2}v_2}}&
\kappa_{12}v_1v_2 &
\kappa_{23}v_2 v_3-A\Gamma_s{\displaystyle\frac{s}{\sqrt{2}}}&
\kappa_{2s}v_2 s-A\Gamma_s{\displaystyle\frac{v_3}{\sqrt{2}}}
\vspace{0.1cm}\\
\kappa_{12}v_1v_2&\kappa_1^2v_1^2&\kappa_{13}v_1v_3&\kappa_{1s}v_1s
\vspace{0.1cm}\\
\kappa_{23}v_2v_3-A\Gamma_s{\displaystyle\frac{s}{\sqrt{2}}}
&\kappa_{13}v_1v_3&\kappa^2_3v_3^2+A\Gamma_s
s{\displaystyle\frac{v_2}{\sqrt{2}v_3}}&\kappa_{3s}v_3s-
A\Gamma_s{\displaystyle\frac{v_2}{\sqrt{2}}}\vspace{0.1cm}\\
\kappa_{2s}v_2 s-A\Gamma_s{\displaystyle\frac{v_3}{\sqrt{2}}}&
\kappa_{1s}v_1s&\kappa_{3s}v_3s-
A\Gamma_s{\displaystyle\frac{v_2}{\sqrt{2}}}&\kappa_{s}^{2}s^2+A\Gamma_s
v_3{\displaystyle\frac{v_2}{\sqrt{2}s}}
\end{array}\right),
\end{eqnarray}
with $\kappa_i^2=G^2/4+{g'}_1^2Q_i^2$,
$\kappa_{1j=2,3}={g'}_1^2Q_1Q_j-G^2/4$, $\kappa_{1s}={g'}_1^2Q_1Q_S$, 
$\kappa_{23}=\Gamma_s^2+{g'}_1^2Q_1Q_j-G^2/4$,
 $\kappa_{js}=\Gamma_s^2+{g'}_1^2Q_jQ_S$, and
$\kappa_{s}^{2}={g'}_1^2Q_S^2$.

In the numerical solution obtained, the values of the
masses of the four scalars are $m_{h^0_{1}}=33$ GeV; $m_{h^0_{2}}=47$ 
GeV; $m_{h^0_{3}}=736$ GeV; $m_{h^0_{4}}=1650$ GeV.

In general supersymmetric models, one of the physical Higgs bosons has
a mass controlled by the electroweak scale from the breaking of $SU(2)_{L}
\times U(1)_{Y}$, while the others may have masses at the scale of the soft
$SUSY$ breaking. In this model, an additional global $U(1)$ 
symmetry is broken when $h_c$ acquires a non-zero VEV, and the mass of the
scalar Higgs associated with this direction is mainly determined by the 
$U(1)$ breaking scale. 
This scale is therefore comparable to the electroweak scale, which indicates
that in the decoupling limit not only one but two Higgs scalars will be
light. In this particular example, the VEV of $h_{c}$ happens to be small
compared with that of the other two Higgs doublets, and hence the lightest
Higgs is mainly associated with $h_c$.
Thus, the lightest Higgs mass satisfies the (tree-level) bound
\begin{eqnarray}
\label{bound1}
m^2_{h^0_{1}}\leq \frac{G^2}{4}v_1^2+g^2_{1'}Q_{1}^2v_1^2=(35 \, {\rm 
GeV})^2,
\end{eqnarray}
obtained from analyzing the potential in the field direction that breaks
the global $U(1)$ symmetry (i.e., the field direction
$h_c$)~\cite{comelli}. 
In the large $s$ limit, the mass saturates this bound.

It is also possible to place a bound on the second-lightest neutral Higgs
scalar~\cite{comelli}
\begin{equation}
\begin{array}{ccc}
m^2_{h^0_{2}} & \leq &
m^2_{h^0_{1}}+\frac{v^2}{v_2^2+v_3^2}[( \frac{G^{2}v_1^2}{4}+ 
g^2_{1'}Q_{1}^2v_1^2-m^2_{h^0_{1}}) ^{\frac{1}{2}}  \\
 & + & 
(\frac{G^2}{4v^2}(v_2^2-v_1^2-v_3^2 ) ^2+ 
\frac{g^2_{1'}}{v^2}(Q_1v_1^2+Q_2v_2^2+Q_3v_3^2)^2     \\
 & + & 2\frac{\Gamma_s^2}{v^2}v_2^2v_3^2
-m^2_{h^0_{1}}) ^{\frac{1}{2}}] ^2= (85 \, {\rm GeV})^2.
\end{array}
\end{equation}
A suitable rotation in field space demonstrates that the second-lightest
Higgs ($h_{2}^{0}$) is basically the real part of the Higgs doublet that
is involved in the $SU(2)$ breaking, while the other two Higgs doublets do
not participate. One of the two rotated doublets ($h_{4}^0$,
$H_{2}^{\pm}$, $A^0$)
consists of the heaviest scalar, pseudoscalar, and charged Higgs, and is
composed mainly of $\bar{h}_c$ and $h_{b}'$, with mass roughly given by
$m_A^0$ (naturally expected to be large in this limit).  The other doublet
($h_{1}^0$, $H_{1}^{\pm}$, massless pseudoscalar) is basically $h_c$, and 
hence the associated fields are light due to the absence of couplings to
the singlet.  The second-heaviest neutral Higgs ($h_{3}^0$) has mass
governed by $M_{Z'}$, and is primarily the singlet.


\section{$P_{2}P_{3}|_{F}$ flat direction}
\label{IV}
\subsection{Effective Superpotential}
The fields involved in the $P_{2}P_{3}|_{F}$ direction are $ \{\varphi_{2},
\varphi_{4}, \varphi_{10}, \varphi_{12}, \varphi_{27}, \varphi_{30} \}$, with VEV's 
\begin{equation}
|\langle \varphi_{27} \rangle|^2 = 2x^2; \ \  
|\langle \varphi_{30} \rangle|^2=2|\langle \varphi_{2} \rangle|^2= 2|\langle
\varphi_{4} \rangle|^2 =2|\langle \varphi_{10} \rangle|^2= 2|\langle
\varphi_{12} \rangle|^2= x^2 \, ,
\end{equation}
where $x=0.013M_{Pl}$, and $\langle \varphi_{10} \rangle$ and $\langle
\varphi_{4} \rangle $ have opposite signs.

The massless states are presented in Table V. 
The effective trilinear couplings for the observable sector states 
are given by  
\begin{eqnarray}
\label{w3ob2}
W_{3} &=& gQ_cu^c_c\bar{h}_c + gQ_cd^c_bh_c +
\frac{\alpha^{(3)}_{5}x^2}{{\sqrt2}M_{Pl}^{2}}Q_ad^c_dh_g 
 + \frac{\alpha^{(4)}_{5}x^2}{{\sqrt{2}}M_{Pl}^{2}}Q_bd^c_dh_g 
  + \frac{\alpha^{(5)}_{5}x^2}{{\sqrt{2}}M_{Pl}^{2}}Q_bd^c_ch_b 
\nonumber \\
&+& \frac{\alpha^{(6)}_{5}x^2}{{\sqrt{2}}M_{Pl}^{2}}Q_ad^c_ch_b
 + \frac{\sqrt{(\alpha^{(1)}_{4})^2+(\alpha^{(1')}_{4})^2}x}{{\sqrt{2}}M_{Pl}} u^{c'}_bd^c_cd^c_d 
+{g\over {\sqrt{2}}}e^c_ah_ah_c + {g\over {\sqrt{2}}}e^c_fh_dh_c \nonumber \\
&+& \frac{\sqrt{(\alpha^{(2)}_{4})^{2}+(\alpha^{(2')}_{4})^{2}}x}{{\sqrt{2}}M_{Pl}}e^{c'}_eh_gh_b
 + \frac{\alpha^{(7)}_{5}x^2}{2M_{Pl}^{2}}e_b^ch_gh_a+
\frac{\alpha^{(8)}_{5}x^2}{2M_{Pl}^{2}}e_i^ch_gh_a + {g\over
{\sqrt{2}}}\bar{h}_ah_c\varphi_{25} + g \bar{h}_ch_b\varphi_{20} 
\nonumber \\
 &+& g\bar{h}_dh_b\varphi_{28} + g\bar{h}_ch_g\varphi_{21} + g\bar{h}_dh_g\varphi_{29}+
\frac{\alpha^{(3)}_{4}\sqrt{2}x}{M_{Pl}}\varphi_{25}\varphi_{21}\varphi_{29}  
+\frac{\alpha^{(3')}_{4}\sqrt{2}x}{M_{Pl}}\varphi_{25}\varphi_{20}\varphi_{28}
\, ,
\end{eqnarray}
in which
\begin{equation}
\begin{array}{ccc}
u_{b}^{c'} &=&\frac{M_{Pl}}{\sqrt{(\alpha^{(1)}_{4} \langle \varphi_{4}
\rangle)^{2}
+(\alpha^{(1')}_{4} \langle \varphi_{12} \rangle)^{2}}}
(\frac{\alpha^{(1)}_{4}}{M_{Pl}}
\langle \varphi_{4} \rangle u^{c}_b
+\frac{\alpha^{(1')}_{4}}{M_{Pl}}
\langle \varphi_{12}\rangle u^c_a); \\
e_{e}^{c'} &=&\frac{M_{Pl}}{\sqrt{(\alpha^{(2)}_{4} \langle \varphi_{12}
\rangle)^{2}
+(\alpha^{(2')}_{4} \langle \varphi_{4} \rangle)^{2}}}
(\frac{\alpha^{(2)}_{4}}{M_{Pl}}
\langle \varphi_{12} \rangle e^{c}_e
+\frac{\alpha^{(2')}_{4}}{M_{Pl}}
\langle \varphi_{4} \rangle e^c_h).
\end{array}
\end{equation}

The superpotential (\ref{w3ob2}) displays the same unrealistic $t-b$ and
$\tau-\mu$ unification, absence of the canonical effective $\mu$ term
(though non-canonical $\mu$ terms are present), and
$L$- violating couplings as in the previous case.      

However, there are new features (see  \cite{cceelw1} for a detailed discussion).
In particular, there are $B$- number violating couplings in the superpotential,
with implications for possible proton decay processes and  $N-\bar
N$-oscillations. There is also a texture in the down-quark sector, with a
possibly realistic $m_s/m_b$ ratio due to the contribution of the original
fifth order operators.

An inspection of Table V shows that there are, in fact, a larger number of
massless states in the observable sector than in the $P_1'P_2'P_3'$
flat direction. Once again, there is some ambiguity in how to identify the
three MSSM lepton doublets (each possible set from the list of massless states
leads to $L$- violating couplings).  However, in this case there is an
additional pair of fields ($\bar{h}_d$, $h_g$) which can play the role of
Higgs doublets.

In addition, $\varphi_{25}$ remains massless at the string scale in this
model.  Therefore, there is a possibility that the $U(1)'$ breaking may
occur along a $D$ flat direction, and hence takes place at an
intermediate scale.  This scenario
requires that the mass-square of the field relevant along the flat direction is
driven negative at a scale much higher than the electroweak scale.
We investigate this possibility in
Section IV.C, and show that it is possible with mild tuning of the soft
supersymmetry breaking parameters at the string scale. On the other hand,
if this condition is not satisfied, the
$U(1)^{'}$ symmetry breaking is naturally at the electroweak scale, coupled to the
breaking of $SU(2)_{L}\times U(1)_{Y}$; we examine this possibility in the
following subsection. 

As in the previous model, we adopt the strategy that the initial boundary
conditions for the hidden sector fields are adjusted to keep their
mass-squares positive at the observable sector symmetry breaking
scale. (In this case, the hidden sector is more involved since
the singlet field $\varphi_{21}$ couples both to the hidden sector fields
and the Higgs doublets (eqn.(27) in \cite{cceelw1}).)


\subsection{Electroweak Scale Symmetry Breaking}


To highlight the unique features of the $P_2P_3|_F$ case as compared with
the
previous example, we choose to study scenarios which
allow for the maximum amount of texture in the quark sector; i.e.,
scenarios in which the Higgs doublets $\bar{h}_{c}$, $h_{c}$,
$h_{g}$ and $h_{b}$, as well as the singlet fields
$\varphi_{20}$ and $\varphi_{21}$, all acquire non-zero VEV's~\footnote{It 
is also possible that $\bar{h}_{a}$, $\bar{h}_{d}$, and the singlet
fields $\varphi_{25}$, $\varphi_{28}$, and $\varphi_{29}$ acquire non-zero 
VEV's, due to the presence of the non-canonical effective $\mu$ terms 
involving these
fields, if the singlet fields involved develop negative mass-squares at the
electroweak scale. With the non-canonical $\mu$ terms $\bar{h}_{a} h_{c}
\varphi_{25}$, 
$\bar{h}_{d} h_{b} \varphi_{28}$ and $\bar{h}_{d} h_{g} \varphi_{29}$ taking
active roles in the symmetry breaking, the massless charginos and neutralinos
(associated with the Higgs doublets that do not have VEV's) as well as the
massless CP odd Higgs scalars (associated with the Higgs doublets that do not
have effective $\mu$ terms), can
be eliminated. However, this implies that the minimum of the potential has
a very complicated structure, which may result in a larger amount of
fine-tuning. We do not investigate such complicated scenarios in this
paper.}. We again restrict
ourselves to scenarios which lead to experimentally allowed $Z^{'}$ masses and
$Z-Z^{'}$ mixing angles.

With these assumptions, an inspection of the resulting scalar potential reveals 
that not all of the phases of the fields which acquire VEV's can be
eliminated by suitable rotations. By redefining the phases
of the fields, the Yukawa coupling and the $A$-parameter associated with
the coupling $\bar{h}_{c} h_{b} \varphi_{20}$ can be taken to be real and
positive.  In the most general case, $\langle
\bar{h}_{c}^{0} \rangle$ and $\langle \varphi_{20} \rangle $ can be chosen
to be real and positive by $SU(2)_{L} \times U(1)_{Y}$ and $U(1)^{'}$
rotations, respectively. Similarly,  $\langle h_{c}^{0} \rangle$ and
$\langle \varphi_{21} \rangle $ can be taken to be real and positive using
global $U(1)$ symmetries of the scalar potential. 
Therefore, there are in general three phases that will
remain non-zero at the minimum of the potential: $\phi_{b}$,
$\phi_{g}$ and $\phi_{A_{2}}$, which are the phases of $\langle h_{b}
\rangle$, $\langle h_{g} \rangle$, and the $A$-parameter associated with
the coupling $\bar{h}_{c} h_{g}\varphi_{21}$. We will take the
$A$-parameter to be real and positive (i.e., we ignore possible explicit
$CP$ violation associated with the soft supersymmetry breaking).
$\phi_{b}$ and $\phi_{g}$ may be non-zero at the minimum of the potential,
leading to spontaneous $CP$-violation and the associated difficulties of
cosmological domain walls. However, $\phi_{b}$ and $\phi_{g}$ vanish at
the minimum for the particular numerical example that we consider. \\

{ (i) Running of the Gauge Couplings and Yukawa Couplings:} \\

We adopt the same strategy for the running of the gauge couplings as the
previous model. The $\beta$ functions for the gauge groups are presented
in Table III. At the electroweak scale, one obtains $g_1=0.40$ (which
includes the factor $k_Y=11/3$) and $g_2=0.46$ ($k_2=1$), yielding
$\sin^2\theta_W=0.17$. 

This model displays a rich set of Yukawa couplings for the observable
sector. The 
initial values of the coupling constants are fixed by string calculations.
In Figure 4, we show the variation of the coupling constants with the
scale.   
\begin{figure}
\vskip -0.3truein
\centerline{
\hbox{
\epsfxsize = 3.5 truein
\epsfbox[70 32 545 740]{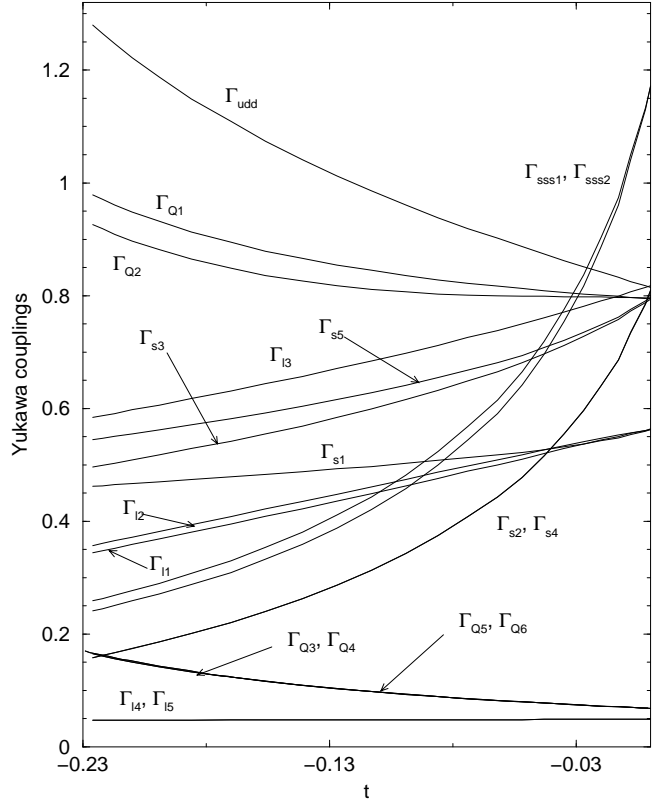}
}
}
\caption{
Running of the Yukawa couplings for the $P_2 P_3 |_{F}$ direction.
The naming and the ordering of the Yukawa coupling constants are the same
as in the previous model. $\Gamma_{udd}$ refers to the
$u^{c'}_bd^c_cd^c_d$ coupling, $\Gamma_{sss1,2}$ refer to the last two
terms in (\ref{w3ob2}).}

\end{figure} 

In addition to the large Yukawa couplings of $\bar{h}_{c}$ and $h_{c}$ to
the top and bottom quarks (which take the values $\Gamma_{Q1}=0.98$ and
$\Gamma_{Q2}=0.93$ at the electroweak scale),
the effective Yukawa couplings from the fifth order (which involve the
other two quark families) have the non-trivial values
$\Gamma_{Q3}=\Gamma_{Q4}=\Gamma_{Q5}=\Gamma_{Q6}=0.17$. 
With the assumption that $h_{g}$ and
$h_{b}$ have non-zero VEV's, these couplings naturally provide a hierarchy
of the quark masses and mixings in the down-type quark sector. 

The $\tau -\mu$ unification is slightly broken, since $\Gamma_{l1}=0.34$
and
$\Gamma_{l2}=0.35$ at the electroweak scale. The Yukawa couplings of the
non-canonical $\mu$ terms are non-trivial as well, with the typical
values of a few times $0.1$
at low energy. Hence, these couplings can be actively involved in the symmetry
breaking, as will be manifest in our numerical example.    \\

{ (ii) Running of the Soft Mass Parameters:} \\

The mass-squares of $h_g$ and $h_b$ tend to remain positive, as their
couplings to quarks arise from fifth order terms in the original 
superpotential and are therefore suppressed. However, the coupling 
$\bar{h}_c h_g \varphi_{21}$ may force $h_g$ to acquire a  non-zero VEV. 
Similarly, $h_{b}$ may acquire a VEV due to the $\bar{h}_c h_b
\varphi_{20}$ coupling.  There are then two distinct electroweak symmetry
breaking  patterns, one in which $h_b$ has a non-zero VEV (i.e., $h_b$ is
naturally  identified as a Higgs field), and  one in which $h_b$ has zero
VEV ($h_b$ could also be identified as a lepton doublet). 

With universal soft mass-squared parameters at the string scale, the
soft mass-squared parameters of $\bar{h}_{c}$ and
$h_{c}$ are driven to negative values, while the mass-squares of all the 
other fields remain positive. This scenario results in a light $Z^{'}$ and
large $Z-Z^{'}$ mixing angle at the electroweak scale, which is excluded
by experiments. We therefore have to consider non-universal boundary conditions
for a realistic solution, in which $h_{g}$, $h_b$, $\varphi_{20}$, and
$\varphi_{21}$ also acquire VEV's. As in the previous example, the effects
of the Yukawa coupling involving hidden sector fields (eqn. (27) of
\cite{cceelw1}) are included in the running Yukawas and soft parameters.

One numerical example we found involves a set of tuned initial
conditions which do not deviate substantially from universality. In Table
VI, we present the numerical values of the parameters for this example, in
which the VEV's of the fields are $\langle \bar{h}_{c}^{0} \rangle=223$
GeV, $\langle h_{c}^{0} \rangle=124$ GeV, $\langle h_{b}^{0} \rangle=17.3$
GeV, $\langle h_{g}^{0} \rangle=24.8$ GeV, 
 $\langle \varphi_{20}\rangle=24.8$ GeV, 
$\langle \varphi_{21} \rangle=4950$ GeV. In
this particular example with $A_{2}$ real and positive,  
$\phi_{b}=\phi_{g}=0$ at the true minimum of the potential, and hence there
is no spontaneous $CP$ violation. For these values,
$M_{Z^{'}}=1.00$ TeV and $\alpha_{Z-Z^{'}}=0.004$. 


We now consider the quark masses and mixings in this model. 
The mass matrix for the down-type quarks is~\cite{cceelw1}
\begin{equation}
M=\left( \begin{array}{ccc}
 \Gamma_{Q3}\langle h_{g}^{0} \rangle  
& \Gamma_{Q6} \langle h_{b}^{0}\rangle
& 0 
\\
 \Gamma_{Q4}\langle h_{g}^{0} \rangle  
& \Gamma_{Q5} \langle h_{b}^{0}\rangle
& 0  
\\
0 & 0 &  \Gamma_{Q2}\langle h_{c}^{0} \rangle 
\end{array} \right) = 
\left( 
\begin{array}{ccc}
4.16 & 3.04 & 0\\
4.16 & 3.04 & 0\\
0    & 0    & 115 
\end{array} \right), 
\end{equation}
after the electroweak scale symmetry breaking. The masses of the down-type
quarks are $m_{d}=0$ GeV,
$m_{s}=7$ GeV, 
and $m_{b}=115$ GeV, where
$d$, $s$ and $b$ stand for the down, strange, and bottom quarks, 
respectively. These are running masses evaluated at $M_{Z}$. Just as in
the example in section III for the $P_{1}^{'}P_{2}^{'}P_{3}^{'}$ 
direction, the scale for $m_{s}$ and $m_{b}$ (as well as for $m_{\tau}$
and $m_{\mu}$) is much too high. Again, we have made no attempt to further
adjust the parameters to obtain a lower $\langle h_{c}^{0} \rangle$.
However, the hierarchy of the relative masses, ($m_{d}:m_{s}:m_{b}$) $\sim$
($0:1:17$) is quite encouraging. This hierarchy, as well as the form of
the
$CKM$ matrix, can be understood from the analytic discussion in
\cite{cceelw1}. 
The $CKM$ matrix, obtained from diagonalizing $M M^{\dagger}$, since there
is no contribution from the up-quark sector, is given by 
\begin{equation}
U^{CKM}=\left(
\begin{array}{ccc}
0.71& 0.71& 0 \\
-0.71& 0.71& 0 \\
0 & 0 & 1.00 \\
\end{array} \right).
\end{equation}   
The maximal mixing between the first two families is not realistic. It 
results from the comparable magnitude of the $\alpha_{5}^{(i)}$ coefficients.
As discussed in \cite{cceelw1}, one can obtain a realistic $m_d/m_s \neq
0$ and a realistic Cabibbo-like mixing by allowing $\alpha_{5}^{(3)}
\neq \alpha_{5}^{(4)}$ and $\alpha_{5}^{(5)}\neq \alpha_{5}^{(6)}$. The
observed relation $\theta_{c} \sim (m_d/m_s)^{1/2}$ ($\sim 0.2$) can be
obtained for specific $\alpha_{5}$'s, but does not hold in general.
$U_{31}^{CKM}=U_{32}^{CKM}=U_{13}^{CKM}=U_{23}^{CKM}=0$ due to the form of
$M$.

We do not present the Higgs, chargino, or neutralino spectra for this
example, as they do not exhibit any qualitatively new features compared to
the $P_{1}^{'}P_{2}^{'}P_{3}^{'}$ direction.

\subsection{Intermediate Scale Symmetry Breaking}

We now investigate the possibility that the $U(1)'$ is broken at an
intermediate scale, along a $D$ flat direction.
This is possible in this model because the field $\varphi_{25}$, which has
a $U(1)'$ charge opposite in sign to the rest of the singlets (with
nonzero $U(1)'$ charges), remains massless at the string scale.  
The viability of this scenario requires that the $D$ flat direction is
also $F$ flat at the trilinear order~\cite{cceel1}; otherwise, the $F$
terms lead to quartic couplings in the scalar potential which force the
VEV's to be at the electroweak scale~\footnote{In principle, it
is possible that total singlets can acquire intermediate scale VEV's,
because the absence of $D$ terms leads to a similar situation as the case
that arises for a  $D$ flat direction.  In this model, the relevant
singlets $\varphi_{28,29}$ also couple to $\varphi_{25}$, which provide
effective $F$ terms that push the VEV's to the electroweak scale, and
hence this scenario is not viable~\cite{cceelw1}.}.
In addition, we require the absence of renormalizable couplings of the
fields in the $D$ flat direction to the Higgs doublet ($\bar{h}_{c}$)
which couples to the top quark, so that there is top quark Yukawa coupling
in the low energy theory.

With these criteria in mind, an inspection of Table V and (\ref{w3ob2})
demonstrates that from the list of singlet fields
($\varphi_{18}-\varphi_{22}$) that have $U(1)'$ charges
equal and opposite to $\varphi_{25}$, the trilinear couplings of $\varphi_{20}$
and $\varphi_{21}$ (to $\bar{h}_{c}$) do not allow for a realistic 
implementation of the 
intermediate scale symmetry breaking.  However, the fields $\{
\varphi_{18}, \varphi_{19},
\varphi_{22} \}$ (which do not have effective trilinear couplings) can be
involved in viable intermediate scale scenarios. 

To achieve the $U(1)'$ breaking, the effective mass-square
$m_{\varphi_{25}}^{2}+m_{\varphi_{18/19/22}}^{2}$ must be driven negative at a
scale higher than the electroweak scale.  
With non-universal boundary conditions, the mass-square of $\varphi_{25}$
can be driven negative, while keeping the mass-squares of $\varphi_{20}$
and $\varphi_{21}$ positive.  The mass-squares of $\varphi_{18,19,22}$ do
not run, due to the absence of Yukawa couplings involving these fields.
Therefore, we can choose initial values of these mass-squares such that 
$m_{\varphi_{25}}^{2}+m_{\varphi_{18/19/22}}^{2}$ is driven negative at an
intermediate scale (while keeping
$m_{\varphi_{25}}^{2}+m_{\varphi_{20/21}}^{2}$
positive). These conditions ensure that the minimum occur along the $D$
flat direction involving $\varphi_{25}$ and $\varphi_{18/19/22}$.   A
range of intermediate scales $\mu_{RAD}$, at which
$m_{\varphi_{25}}^{2}+m_{\varphi_{18/19/22}}^{2}$ crosses zero, 
can be obtained by adjusting the initial values of the parameters. In
Figure 5 we display the scale variation of the mass-squares with initial
values that lead to $\mu_{RAD} \sim 10^{12}$ GeV.  This example requires
nonuniversal initial values, with $m^2_{\varphi_{20,21}}$ larger by
factors of around 2 to 9 than the others.

\begin{figure}
\vskip -0.3truein
\centerline{
\hbox{
\epsfxsize = 2.8 truein
\epsfbox[70 32 545 740]{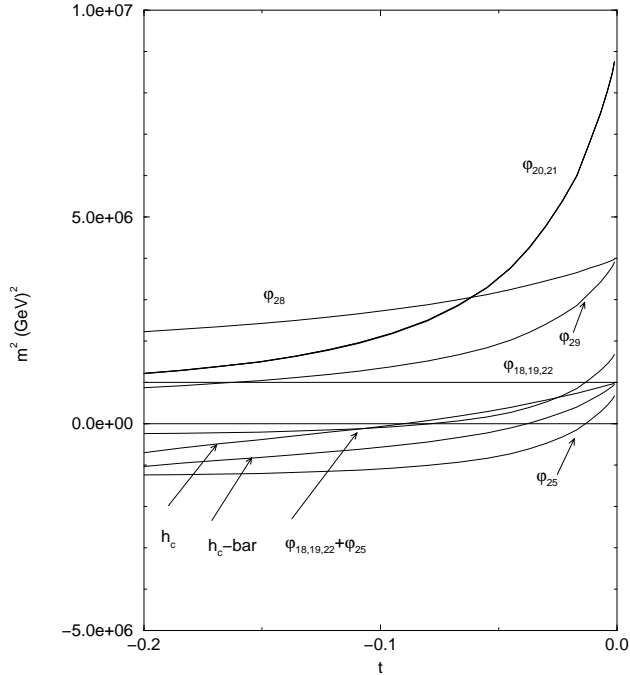}
}
}
\caption{
Running of the relevant mass-squared parameters for the $P_2 P_3 |_{F}$
flat direction, for the $U(1)'$ symmetry breaking scenario along the $D$ 
flat direction $\varphi_{25}+\varphi_{18/19/22}$, with $\mu_{RAD} \sim
10^{12}$ GeV. }

\end{figure}

In the intermediate scale symmetry breaking scenario~\cite{cceel1}, the
potential can be stabilized by radiative corrections, or by
nonrenormalizable self-couplings of the singlet fields in the flat
direction; i.e., terms of the form
$(\varphi_{25}\varphi_{18/19/22})^n/M^{2n-3}$.
Such nonrenormalizable self-couplings have a number of
sources.  For example, they can be present in the original superpotential
or induced by vacuum restabilization from higher-dimensional
operators.  In addition, these terms can arise from the decoupling of
the heavy fields~\cite{ceew}, as well as from a nonminimal K\"{a}hler
potential (either from the original couplings from the effective string 
theory or could be induced either after vacuum restabilization or due to 
decoupling effects). 

Within our assumption of a minimal K\"{a}hler potential, we have checked
to determine if these nonrenormalizable self-couplings are in fact present 
 and found that these couplings do not survive the string
selection rules~\cite{cceelw1}.  
To determine if such terms can be induced at the leading order in 
the superpotential from decoupling~\cite{ceew}, it
is necessary to check that there are no trilinear terms in the effective
superpotential that involve two powers of the fields in the intermediate
scale flat direction and one power of a heavy field.  In this case, gauge
invariance restricts these terms to be of the type $\varphi_{25}
\varphi_{18/19/22}\varphi$, in which $\varphi$ is a heavy gauge singlet
under $U(1)_Y$ and $U(1)'$ ($\varphi$ can be one of the fields
in the $P_2P_3|_F$ flat direction). We find that in this example there
are no such terms.  
Therefore, the symmetry breaking is purely radiative in origin, and the
VEV's are very close to the scale $\mu_{RAD}$. 

The next step is to investigate the electroweak 
symmetry breaking after the intermediate scale $U(1)'$ breaking.  The
electroweak symmetry breaking has different features than in
the previous case.  For example, the field $h_c$, which played an
important role in the electroweak symmetry breaking due to its Yukawa
couplings to the bottom quark and the $\mu$ and $\tau$ leptons, acquires
an intermediate scale mass and decouples from the theory, which seems to
leave these MSSM fields massless. However, it may be possible
that effective mass for the exotics and effective Yukawa couplings for the
MSSM fields are generated from nonrenormalizable operators involving the
fields in the intermediate scale \cite{cceel1}. Similarly, effective $\mu$ terms may be
generated by NRO's.  In principle, the determination of the
complete set of such NRO's is a necessary first step in the analysis,
since the renormalization group equations are affected when the
fields with intermediate scale masses decouple.  
The determination of the complete set of such operators and the subsequent
electroweak symmetry breaking patterns is beyond the scope of this paper
(but is the subject of a future investigation).

\section{Conclusions}
\label{V}
This paper is the culmination of the program that sets out to derive 
the phenomenological implications of a class of quasi-realistic string
models. We have determined the observable sector gauge  symmetry breaking
patterns and the mass spectrum for a class of representative string-scale
 flat directions of a prototype model by merging the top-down
approach (employing the string results for the effective superpotential
couplings) and the bottom-up approach (adding the soft supersymmetry
breaking mass parameters by hand).   

The set of top-down inputs  builds on the results of our previous work:
(i) the classification of
the $D$ flat directions of the model that cancel the anomalous $D$ term
and can be proven to be $F$ flat to all orders~\cite{cceel2}, and (ii)
the determination of the effective theory along such flat
directions. In particular, we have utilized the results
of~\cite{cceelw1}, in which the complete mass spectrum and effective
trilinear couplings of the observable sector were presented for two
representative flat directions of Model 5 of \cite{chl} (CHL5), 
which leave an additional $U(1)'$
as well as the SM gauge group unbroken. The first representative flat
direction has a minimal number of couplings of the observable sector
fields, while the second flat direction has a richer structure of
couplings, with implications for fermion textures.

The mass spectrum and couplings of these effective theories are
not realistic, as expected.  We found that in general there are a number
of superfields which remain massless at the string scale, and that while
the scalars acquire masses from the soft supersymmetry breaking terms,
within our assumptions there is no mechanism for some of the fermions to
acquire masses.
The gauge coupling unification is not realistic (although better than
expected due to the amount of additional matter superfields and the higher
Ka\v c-Moody level in the model), and the hidden sector gauge groups 
are not asymptotically free, thus disallowing an implementation of
dynamical supersymmetry breaking scenarios. 
In addition, the effective trilinear couplings have a number of
nonstandard features which were examined in detail in~\cite{cceelw1}, such
as the absence of a canonical effective $\mu$ term, presence of baryon
and lepton number violating couplings, as well as (potentially realistic)
hierarchies of  fermion masses.

Since the purpose of this program is to explore the general features 
of this class of quasi-realistic models systematically, and not to search
for a specific, fully realistic model (an unlikely possibility), 
we continue the  analysis by investigating the gauge symmetry breaking
patterns and the low energy spectrum for  the representative  examples.  
However, this study requires  the implementation of supersymmetry
breaking, which we parameterize by soft supersymmetry breaking
masses put in by hand at the string scale, due to the absence of a
satisfactory scenario for supersymmetry breaking in string theory. 
This introduces free parameters  in the  effective theory, and thus the
unique predictive power of a particular string vacuum is lost.  
In particular, the concrete results for the low
energy spectrum depend on the initial conditions for the soft masses at the
string scale.

We chose to analyze
scenarios which lead to a realistic $Z-Z'$ hierarchy; as argued on general
grounds in~\cite{cl,cdeel,lw}, the breaking scale of the $U(1)'$ is at the
electroweak scale, or at an intermediate scale (if the symmetry breaking
occurs along a $D$ flat direction).
For each representative flat direction we determine the low energy spectrum 
explicitly  for a typical choice of initial conditions
 that yield a realistic $Z-Z'$ hierarchy. The  emphasis was on the
study of the Higgs  sector, in order to contrast
its features with that encountered in both the MSSM and string
motivated models with an additional $U(1)'$~\cite{cdeel,lw}. 
 In the  symmetry breaking scenarios for these two representative
examples, the novel feature is that the number of 
 Higgs fields that participate in the symmetry breaking is larger
 than that  assumed in \cite{cdeel} (there are at least three Higgs
doublets and one SM singlet participating in the symmetry breaking
process). In addition, the presence of some and absence of other
non-canonical $\mu$ terms implies new patterns in the low energy mass
spectrum.

For the first representative flat direction, the symmetry breaking scale
of the $U(1)'$ is at the electroweak scale, because all of the $U(1)'$
charged singlets that remain massless at the string scale have charges of
the same sign. To obtain a realistic $Z-Z'$ hierarchy, we found it is
necessary to have a nonminimal Higgs sector, with three Higgs doublets and
one singlet (which has a large VEV).  This scenario can be obtained with
mildly nonuniversal soft supersymmetry breaking mass parameters at the
string scale.  We present the complete mass spectrum, including
that of  the supersymmetric partners, for a typical choice of initial
conditions. The resulting 
mass spectrum at the electroweak scale includes massless and ultralight 
charginos, neutralinos, and Higgs bosons, due to the absence of enough
canonical or non-canonical effective $\mu$ terms. In particular, the
absence of an effective $\mu$ term involving one of the Higgs doublets
provides an additional global $U(1)$ symmetry in the scalar potential that
is spontaneously broken, resulting in a Goldstone boson present in the low
energy spectrum.   In addition, the mass of lightest neutral Higgs boson
is controlled by the scale of the breakdown of this global symmetry, and
thus obeys a different bound than the traditional bound in the MSSM.

For the second representative flat direction, the $U(1)'$ breaking can be
either at the electroweak scale or at an intermediate scale.  Once again,
the electroweak scale scenario requires an extended Higgs sector to obtain
a $Z'$ that is consistent with experiment.  
In addition, to study the fermion texture in the down-quark sector of this
model, additional Higgs doublets are needed to acquire VEV's.  In contrast
to the previous model, there may be CP-violating phases in the Higgs sector
of the model, though these are absent for the specific example considered.   
The electroweak scale symmetry breaking scenario is
achievable with mildly nonuniversal boundary conditions. 
There is a fairly realistic $d:s:b$ mass hierarchy, although the 
absolute scale is much too high. 
The corresponding $CKM$ matrix has no mixings of the first two families
with the third, and an unrealistic maximal 
mixing between the first and second generations.   

The intermediate scale symmetry breaking scenario can be achieved with
mild nonuniversality of the soft supersymmetry breaking mass parameters,
at a range of intermediate scales.  It is purely
radiative in origin,  because of the absence of the relevant
nonrenormalizable terms (and trilinear terms involving the heavy fields) 
in the superpotential.  
We plan to address the electroweak symmetry breaking in this scenario 
after decoupling the heavy fields, which requires a detailed analysis of a
class of nonrenormalizable operators, in a future paper.

Although the analysis of the low energy implications for
the representative examples studied in the paper reveals a number 
of  additional unacceptable  phenomenological 
consequences, it nevertheless demonstrates new features of the gauge  
symmetry breaking patterns, in particular that associated with the Higgs 
sector of the theory. The type of non-minimal
extensions of the Higgs sector and their specific couplings, which we 
encountered in the analysis of the specific string models  should be 
of general interest in the phenomenological investigation of models 
beyond the MSSM, and thus deserves further study.

\acknowledgments
We would like to thank J. Lykken for making available to us the
program that generates the massless spectrum of free fermionic
string vacua as well as the original program that calculates the
superpotential couplings.   
L.E. acknowledges G. Kane for useful discussions. This work
was supported in part by U.S. Department of Energy Grant No.
DOE-EY-76-02-3071. 
\newpage

\def\B#1#2#3{\/ {\bf B#1} (19#2) #3}
\def\NPB#1#2#3{{\it Nucl.\ Phys.}\/ {\bf B#1} (19#2) #3}
\def\PLB#1#2#3{{\it Phys.\ Lett.}\/ {\bf B#1} (19#2) #3}
\def\PRD#1#2#3{{\it Phys.\ Rev.}\/ {\bf D#1} (19#2) #3}
\def\PRL#1#2#3{{\it Phys.\ Rev.\ Lett.}\/ {\bf #1} (19#2) #3}
\def\PRT#1#2#3{{\it Phys.\ Rep.}\/ {\bf#1} (19#2) #3}
\def\MODA#1#2#3{{\it Mod.\ Phys.\ Lett.}\/ {\bf A#1} (19#2) #3}
\def\IJMP#1#2#3{{\it Int.\ J.\ Mod.\ Phys.}\/ {\bf A#1} (19#2) #3}
\def\nuvc#1#2#3{{\it Nuovo Cimento}\/ {\bf #1A} (#2) #3}
\def\RPP#1#2#3{{\it Rept.\ Prog.\ Phys.}\/ {\bf #1} (19#2) #3}
\def\etal{{\it et al\/}}

\bibliographystyle{unsrt}

\begin{references}
%

\bibitem{cceel2}{G. Cleaver, M. Cveti\v c, J.R. Espinosa,
L. Everett, and P. Langacker, \NPB{525}{98} 3.} 
\bibitem{cceel3} {G. Cleaver, M. Cveti\v c, J.R. Espinosa,
L. Everett, and P. Langacker, [hep-th/9805133], submitted to {\it
Nucl.\ Phys.\ }{\bf B}.}
\bibitem{cceelw1}{G. Cleaver, M. Cveti\v c, J.R. Espinosa, L. Everett, P.
Langacker and  J. Wang  ,[hep-ph/9807479].}

\bibitem{orbifolds} 
L.~Ib\'a\~nez, J.E.~Kim, H.P.~Nilles and F.~Quevedo, \PLB{191}{87}{282};
J.A.~Casas and C.~Mu\~noz, \PLB{209}{88}{214} and \B{214}{88}{157}; 
J.A.~Casas, E.~Katehou and C.~Mu\~noz, \NPB{317}{89}{171}; 
A.~Font, L.~Ib\'a\~nez, H.P.~Nilles and F.~Quevedo,
\PLB{210}{88}{101};
A.~Chamseddine and M.~Quir\'os, \PLB{212}{88}{343},
\NPB{316}{89}{101};
A.~Font, L.~Ib\'a\~nez, F.~Quevedo and A.~Sierra,
\NPB{331}{90}{421}.
%
\bibitem{calabiyau} B.~Greene, K.~Kirlin, P.~Miron and G.G.~Ross,
\NPB{278}{86}{667} and \B{292}{87}{606}


\bibitem{NAHE} 
I. Antoniadis, J. Ellis, J. Hagelin, and D.V. Nanopoulos,
 \PLB{231}{1989}{65};  I. Antoniadis, G.K.
Leontaris and  J. Rizos, {\it Phys. Lett.} {\bf B245} (1990) 161;
A.E. Faraggi, \NPB{387}{92}{239};
  [hep-th/9708112]; A.E. Faraggi and D.V. Nanopoulos, \PRD{48}{93}{3288}.

\bibitem{FNY1}{A. Faraggi, D.V. Nanopoulos, and K. Yuan, \NPB{335}{90}{347};
A. Faraggi, \PRD{46}{92}{3204}.} 

\bibitem{AF1}{A. Faraggi, \PLB{278}{92}{131}, 
\NPB{403}{93}{101} and \PLB{339}{94}{223}.} 

\bibitem{chl}{S. Chaudhuri,  G. Hockney and J. Lykken, \NPB{469}{96}{357}.}

\bibitem{dienes} K.R. Dienes, Phys. Rept. {\bf 287}, 447 (1997).

\bibitem{CEW}{M. Cveti\v c,  L. Everett, and J. Wang, [hep-ph/9808321].}

\bibitem{ceew}{M. Cveti\v c,  L. Everett, and J. Wang, [hep-ph/9807321], 
to be published in {\it Nucl. Phys.} {\bf B}.}


%
\bibitem{C}{L.~Dixon and M.~Cveti\v c, unpublished; M.~Cveti\v c, 
\PRL{59}{87}{2829}, 
D. Bailin, D. Dunbar, and A. Love,
\PLB{219}{89}{76};
S. Kalara, J. L\'opez, and D.V. Nanopoulos,  
\PLB{245}{90}{421}, \NPB{353}{91}{650};
A. Faraggi, \NPB{487}{97}{55}.}







%

%
%
\bibitem{cl}{M.~Cveti\v c and P.~Langacker,
\PRD{54}{96}{3570}, \MODA{11}{96}{1247} and [hep-ph/9707451].}
%
 \bibitem{cdeel}{M.~Cveti\v c, D.A.~Demir, J.R.~Espinosa, L.~Everett and 
P.~Langacker, \PRD{56}{97}{2861}. For related work, see Ref~\cite{cceel1,SY,lw,KM}.
}
\bibitem{cceel1}{G. Cleaver, M. Cveti\v c, J.R.
Espinosa, L. Everett, and  P. Langacker, {\it Phys. Rev.} {\bf 
D57}, 2701 (1998).}
%
\bibitem{SY}{ D. Suematsu and Y. Yamagishi, Int. J. Mod. Phys {\bf A10} 
(1995) 4521.} 
\bibitem{lw}{P. Langacker and J. Wang,
[hep-ph/9804428], to be published in 
{\it Phys. Rev.} {\bf D}.}
\bibitem{KM}{E. Keith and E. Ma,  {\it Phys. Rev.} {\bf D56} (1997) 7155.}

\bibitem{Kap}{V. Kaplunovsky, \NPB{307}{88}{145} and  Erratum-ibid. {\bf
B382} (1992)436.}

\bibitem{C}{L.~Dixon and M.~Cveti\v c, unpublished; M.~Cveti\v c,
\PRL{59}{87}{2829}, D. Bailin, D. Dunbar, and A. Love,
\PLB{219}{89}{76};
S. Kalara, J. L\'opez, and D.V. Nanopoulos,
\PLB{245}{90}{421}, \NPB{353}{91}{650};
A. Faraggi, \NPB{487}{97}{55}.}

%
%
%
%
\bibitem{mbmtau}{P. Langacker and N. Polonsky, {\it Phys. Rev.} {\bf D49}
(1994) 1454;
M. Carena {\it et al.}, {\it Nucl. Phys.} {\bf B426} (1994) 269;
V. Barger, M.S. Berger and P. Ohmann, {\it Phys. Rev.} {\bf D47} (1993)
1093.}

\bibitem{comelli}{D. Comelli and J. R. Espinosa, \PLB{388}{96}{793}; J.
R. Espinosa, Lectures presented at the XXIV ITEP Winter School, Snegiri
(Russia), February 1996, [hep-ph/9606316].}
\end{references}

\newpage

\begin{center}
\begin{tabular}{|c|c|c|c|}
\hline\hline
$(SU(3)_C,SU(2)_L,$& &
    $6Q_{Y}$&$100Q_{Y'}$\\
$SU(4)_2,SU(2)_2)$ &&      &\\
\hline\hline
(3,2,1,1):&$Q_a$& 1&68\\
&$Q_b$ &  1&68\\
&$Q_c$ & 1&$-$71\\
\hline
($\bar{3}$,1,1,1):&$u^c_a$&  $-$4&6\\
&$u^c_b$ & $-$4&6\\
&$u^c_c$ & $-$4&$-$133\\
&$d^c_a$ & 2&$-$3\\
&$d^c_b$ & 2&136\\
&$d^c_c$ & 2&$-$3\\
&$d^c_d$ & 2&$-$3\\
\hline
(1,2,1,1):&$\bar{h}_a$ &  3&$-$74\\
&$\bar{h}_b$ & 3&65\\
&$\bar{h}_c$ & 3&204\\
&$\bar{h}_d$ &  3&65\\
&$h_a$ &  $-$3&74\\
&$h_b$ & $-$3&$-$65\\
&$h_c$ &  $-$3&$-$65\\
&$h_d$ & $-$3&$-$65\\
&$h_e$ & $-$3&$-$204\\
&$h_f$ & $-$3&$-$65\\
&$h_g$ & $-$3&$-$65\\
\hline
(3,1,1,1):&${\cal D}_a$& $-$2&$-$136\\
\hline
\hline
\end{tabular}
\end{center}
\noindent Table Ia: List of non-Abelian non-singlet observable sector
fields in the model with their charges under hypercharge and $U(1)'$.

\newpage
\begin{center}
\begin{tabular}{|c|c|c|c|}
\hline\hline
$(SU(3)_C,SU(2)_L,$&
    
&$6Q_{Y}$&$100Q_{Y'}$\\
$SU(4)_2,SU(2)_2)$ &      &&\\
\hline\hline
(1,2,1,2):& $D_{1-4}$&  0&0\\
\hline
(1,1,4,1):& $F_{1,2}$&  $-$3&$-$65\\
\hline
(1,1,$\bar{4}$,1):& $\bar{F}_{1,2}$&  3&65\\
 &$\bar{F}_{3-6}$&  3&65\\
 &$\bar{F}_{7,8}$&  $-$3&$-$65\\
 &$\bar{F}_{9,10}$&  $-$3&$-$65\\
\hline
(1,1,1,2):& $H_{1,2}$&  3&65\\
 &$H_{3,4}$&  3&204\\
 &$H_{5,7}$&  $-$3&$-$65\\
 &$H_{6,8}$& $-$3&74\\
\hline
(1,1,4,2):& $E_{1,2}$& 0&$-$139\\ 
 &$E_{3}$&0&0\\
 &$E_{4,5}$&0&0\\
\hline
(1,1,$\bar{4}$,2):& $\bar{E}_1$&0&0\\
\hline
(1,1,6,1):& $S_1$&0&0\\
 &$S_2$&0&0\\
 &$S_3$&0&0\\
 &$S_4$&0&0\\
 &$S_5$&0&0\\
 &$S_{6,7}$&0&139\\
 &$S_8$&0&0\\
\hline
(1,1,1,3): &$T_1$&0&0\\
 &$T_2$&0&139\\
 &$T_3$&0&0\\
\hline
\hline
\end{tabular}
\end{center}
\noindent Table Ib: List of non-Abelian 
non-singlet hidden sector fields in the
model with their charges under hypercharge and $U(1)'$.

\newpage
\begin{center}
\begin{tabular}{|c|c|c||c|c|c|}
\hline\hline
 &  $6Q_{Y}$&$100Q_{Y'}$ & & $6Q_{Y}$ &$100Q_{Y'}$ \\
 &          &            & &          & \\
\hline\hline
$e^c_{a,c}$ &6&$-$9 & $e^c_b$  &6&$-$9\\
$e^c_{d,g}$ &6&130 & $e^c_e$ &6&130\\
$e^c_f$ &6&130 & $e^c_h$ &6&130\\
$e^c_i$ &6&$-$9 & $e_{a,b}$ &6&$-$130\\
$e_c$ &6&$-$130 & $e_{d,e}$ &$-$6&9\\
$e_f$ &$-$6&$-$269 & & & \\
\hline
\hline
 &  $6Q_{Y}$&$100Q_{Y'}$ & & $6Q_{Y}$ &$100Q_{Y'}$ \\
 &          &            & &          & \\
$\varphi_{1}$ &0&0 & $\varphi_{2,3}$ &0&0 \\
$\varphi_{4,5}$ &0&0 & $\varphi_{6,7}$ &0&0 \\
$\varphi_{8,9}$ &0&0 & $\varphi_{10,11}$&0&0 \\
$\varphi_{12,13}$&0&0 & $\varphi_{14,15}$&0&0\\
$\varphi_{16}$ &0&0 & $\varphi_{17}$ &0&0 \\
$\varphi_{18,19}$&0&$-$139 & $\varphi_{20,21}$&0&$-$139 \\
$\varphi_{22}$ &0&$-$139 & $\varphi_{23}$ &0&0 \\
$\varphi_{24}$ &0&0 & $\varphi_{25}$ &0&139 \\
$\varphi_{26}$ &0&0 & $\varphi_{27}$ &0&0 \\
$\varphi_{28,29}$&0&0 & $\varphi_{30}$ &0&0 \\
\hline
\hline
\end{tabular}
\end{center}
\noindent Table Ic: List of non-Abelian singlet fields in
the model with their charges under hypercharge and $U(1)'$.

\newpage
\begin{center}
\begin{tabular}{|c|}
\hline\hline
Massless Fields\\  
\hline\hline
 $Q_{a}$,$Q_{b}$,$Q_{c}$\\
$u^c_{a}$,$u^c_{b}$,$u^c_{c}$\\
$d^c_{a}$,$d^c_{b}$,$d^c_{c}$,$d^c_{d}$,${\cal D}_a$\\
$\bar{h}_a$,$\bar{h}_c$\\
$h_a$,$h_c$,$h_d$,$h_e$\\
$h_b'=\frac{1}{\sqrt{N_1}}[-\frac{\alpha^{(1)}_4
|\psi_2|^2+\alpha^{(2)}_4 (|\psi_1|^2-|\psi_2|^2)}{M_{Pl}}
h_f+\sqrt{2}gx h_b$]\\
$e^c_a,e^c_b,e^c_c,e^c_e,e^c_f,e^c_h,e^c_i$\\
$e_c,e_d,e_e,e_f$\\
$\varphi_{3},\varphi_{11},$
$\varphi_{18},\varphi_{19},\varphi_{20},\varphi_{22},\varphi_{24},\varphi_{28}
$\\
$\varphi_{12}'=\frac{1}{|\psi_1|}(-\sqrt{|\psi_1|^2-|\psi_2|^2}\varphi_{12} 
+|\psi_2|\varphi_4)$\\
$D_1,D_2,D_3,D_4$\\
$\bar{F}_3,\bar{F}_4,\bar{F}_5,\bar{F}_6,\bar{F}_7,
\bar{F}_8,\bar{F}_9,\bar{F}_{10}$\\
$H_{1},H_{2},H_{3},H_{4},H_{5},H_{6},H_{7},H_{8}$\\
$E_1,E_2,E_3,E_4,E_5, \bar{E}_1$\\
$S_{2},S_{4},S_{6},S_{7},S_8$\\
$S_1'=\frac{1}{|\psi_1|}(-\sqrt{|\psi_1|^2-|\psi_2|^2}S_3+|\psi_2|S_1)$\\$T_{1},T_{2},T_{3}$\\
\hline
\hline
\end{tabular}
\end{center}
\noindent Table II: List of massless states (excluding the two moduli) 
for the $P_1'P_2'P_3'$
flat direction~\cite{cceelw1}, where $N_1\equiv 
2g^2x^2+(\alpha^{(1)}_4|\psi_2|^2/M_{Pl})^2$  with the VEV
parameters $x, \  \psi_1$ and $\psi_2$ defined in eqs. (\ref{vevs}).

\begin{center}
\begin{tabular}{|c|c|c|c|c|c|c|c|}
\hline\hline

 {\rm Effective} $\beta$ & $\beta_1$ & $\beta_2$ & $\beta_3$ &
$\beta_{1^{'}}$ & $\beta_{11^{'}}$ & $\beta_{2hid}$ & $\beta_{4hid}$ \\
\hline \hline
$P_1^{'}P_2^{'}P_3^{'}$ Flat Direction & $10.0$ & $6.0$ & $-2.0$ & $10.2$
& $4.8$ & $10.0$ & $2.0$ \\
$P_2P_3|_F$ Flat Direction & $10.3$ & $7.0$ & $-2.0$ & $10.6$ & $5.0$ &
$10.0$ & $3.0$ \\
\hline \hline
\end{tabular}
\end{center}
\noindent Table III: Effective beta-functions for the two representative
flat directions, defined as  $\beta_{i}\equiv
\beta^0_i/k_i$, where  $\beta^0_i$  and $k_i$ are the beta-function
and the
Ka\v c-Moody level for  a particular gauge group factor, respectively.
 The subscripts $1,\ 2, \ 3, \ 1', \  {2hid},
\ {4hid}$  refer to $U(1)_Y, \ SU(2)_L, \ SU(3)_C, \ U(1)',\ SU(2)_2, \
SU(4)_2$ gauge group factors and $11'$ refers to the $U(1)_Y$ and $U(1)'$
kinetic  mixing. The Ka\v c-Moody levels are $k_1=11/3$, $k_2=k_3=1$,
$k_{1'}\simeq
16.67$,  and  $k_{2hid}=k_{4hid}=2$.

\newpage 

\begin{center}
\begin{tabular}{||c||c|c||c||c|c||}
\hline
\hline

 &  $M_Z$ & $M_{String}$ & & $M_Z$ & $M_{String}$  \\ \hline
$g_1$& 0.41& 0.80&$M_1$& 444&1695 \\
$g_2$& 0.48& 0.80&$M_2$& 619& 1695\\
$g_3$& 1.23& 0.80&$M_3$& 4040& 1695\\
$g_1'$& 0.43& 0.80&$M_1'$& 392& 1695\\
$\Gamma_{Q1}$&0.96 & 0.80&$A_{Q1}$ & 3664&8682\\
$\Gamma_{Q2}$&0.93 & 0.80& $A_{Q2}$&4070&9000\\
$\gamma_{Q3}$&0.27 & 0.08& $A_{Q3}$&5018&1837\\
$\Gamma_{l1}$& 0.30& 0.56&$A_{l1}$ &$-$946&4703\\
$\Gamma_{l2}$& 0.36& 0.56&$A_{l2}$ &$-$707&4532\\
$\Gamma_{l3}$& 0.06& 0.05&$A_{l3}$ &4613&4425\\
$\Gamma_{l4}$& 0.11& 0.13&$A_{l4}$ &4590&4481\\
$\Gamma_{s}$& 0.22& 0.80 &$A$      &1695&12544\\
$m_{Q_c}^2$ & $(2706)^2$ & $(2450)^2$ & 
$m_{d_d}^2$ & $(4693)^2$ & $(2125)^2$ \\ 
$m_{u_c}^2$ & $(2649)^2$ & $(2418)^2$ & 
$m_{d_c}^2$ & $(2734)^2$ & $(2486)^2$\\ 
$m_{\bar{h}_c}^2$ & $(1008)^2$ & $(5622)^2$ &
$m_{h_b'}^2$&$(826)^2$&$(2595)^2$\\ 
$m_{\varphi_{20'}}^2$ & $-(518)^2$ & $(6890)^2$ &
$m_{\varphi_{22'}}^2$ &$(3031)^2$&$(11540)^2$\\ 
$m_{h_a}^2$ & $(3626)^2$ & $(3982)^2$ & 
$m_{h_c}^2$ & $-(224)^2$&$(5633)^2$\\ 
$m_{h_d}^2$ & $(3666)^2$ & $(4100)^2$ & 
$m_{h_e}^2$ & $(4274)^2$ & $(4246)^2$\\ 
$m_{e_a}^2$ & $(2770)^2$ & $(3564)^2$ & 
$m_{e_f}^2$ & $(2780)^2$ & $(3958)^2$\\ 
$m_{e_e}^2$ & $(4195)^2$ & $(4254)^2$ & 
$m_{e_h}^2$ & $(4259)^2$ & $(4236)^2$\\
\hline\hline
\end{tabular}
\end{center}
\noindent Table IV: $P_1'P_2'P_3'$ flat direction: values of the parameters at $M_{String}$ and $M_{Z}$, with $M_{Z'}=735$
GeV and $\alpha_{Z-Z'}=0.005$.  All mass parameters are given in GeV. Other soft supersymmetry breaking mass parameters approximately decouple, and are not presented.

\newpage
\begin{center}
\begin{tabular}{|c|}
\hline\hline
Massless Fields\\  
\hline\hline
 $Q_{a}$,$Q_{b}$,$Q_{c}$\\
$u^c_{a}$,$u^{c}_{b}$,$u^c_{c}$\\
$d^c_{a}$,$d^c_{b}$,$d^c_{c}$,$d^c_{d}$,${\cal D}_a$\\
$\bar{h}_a$,$\bar{h}_c$, $\bar{h}_d$ \\
$h_a$,$h_b$,$h_c$,$h_d$,$h_e$, $h_g$ \\
$e_a^c$, $e_b^c$, $e_c^c$, $e_e^c$, $e_f^c$, $e_h^c$, $e_i^c$ \\
$e_c$, $e_d$, $e_e$, $e_f$ \\
$\varphi_{5}, \varphi_{6}, \varphi_{13}, \varphi_{14},
\varphi_{18}, \varphi_{19}, \varphi_{20}, \varphi_{21}, \varphi_{22}, 
\varphi_{23}, \varphi_{24}, \varphi_{25},\varphi_{28},
\varphi_{29}$ \\
$\varphi_{11'}=(\varphi_{3}-\varphi_{11})/\sqrt{2}$, \\
$\varphi_{17'}=(\frac{gx}{{2}}\varphi_{17}+\frac{\alpha^{(2)}_{5}}{M_{Pl}^{2}}
\frac{x^{3}}{2}\varphi_{16})/\sqrt{N_2}$\\
$\varphi_{26'}=(-\frac{gx}{{2}}\varphi_{26}-\frac{\alpha^{(1)}_{5}}
{M_{Pl}^{2}}\frac{x^{3}}{2}\varphi_{1})/\sqrt{N_1}$\\
$D_1,D_2,D_3,D_4$\\
$\bar{F}_3,\bar{F}_4,\bar{F}_5,\bar{F}_6,\bar{F}_7,
\bar{F}_8,\bar{F}_9,\bar{F}_{10}$\\
$H_{1},H_{2},H_{3},H_{4},H_{5},H_{6},H_{7},H_{8}$\\
$E_1,E_2,E_3,E_4,E_5, \bar{E}_1$\\
$S_1,S_{2},S_3,S_{4},S_5,S_{6},S_{7},S_8$\\
$T_{1},T_{2},T_{3}$\\
\hline\hline
\end{tabular}
\end{center}
\noindent Table V: List of massless states  for the $P_2P_3|_F$
flat direction, where $N_{1/2}\equiv
g^2x^{2}/4+(\alpha_5^{(1/2)}x^3/2M_{Pl}^{2})^2$.

\newpage 

\begin{center}
\begin{tabular}{||c||c|c||c||c|c||}
\hline
\hline

 &  $M_Z$ & $M_{String}$ & & $M_Z$ & $M_{String}$  \\ \hline
$g_1$& 0.40& 0.80&$M_1$& 251& 1000 \\
$g_2$& 0.46& 0.80&$M_2$& 330& 1000\\
$g_3$& 1.23& 0.80&$M_3$& 2380& 1000\\
$g_1'$& 0.42& 0.80&$M_1'$& 225& 1000\\
$\Gamma_{Q1}$&0.98 & 0.80& $A_{Q1}$ & 828 &702\\
$\Gamma_{Q2}$&0.93 & 0.80& $A_{Q2}$ &2120 &1227\\
$\Gamma_{Q3}$&0.17 & 0.07& $A_{Q2}$ &3290 &1510\\
$\Gamma_{Q4}$&0.17 & 0.07& $A_{Q2}$ &3290 &1510\\
$\Gamma_{Q5}$&0.17 & 0.07& $A_{Q2}$ &4570 &1308\\
$\Gamma_{Q6}$&0.17 & 0.07& $A_{Q2}$ &4570 &1308\\
$\Gamma_{udd}$& 1.29& 0.82& $A_{udd}$ &2910 &3239\\
$\Gamma_{l1}$& 0.34& 0.56&$A_{l1}$ &$-$240&1234\\
$\Gamma_{l2}$& 0.35& 0.56&$A_{l2}$ &$-$197&1234\\
$\Gamma_{l3}$& 0.58& 0.82&$A_{l3}$ &419&1659\\
$\Gamma_{l4}$& 0.05& 0.05&$A_{l4}$ &807&1351\\
$\Gamma_{l5}$& 0.05& 0.05&$A_{l4}$ &807&1351\\
$\Gamma_{s1}$& 0.46& 0.56&$A_{s1}$ &893&1381\\
$\Gamma_{s2}$& 0.16& 0.80&$A_{1}$  &350&3336\\
$\Gamma_{s3}$& 0.49& 0.80&$A_{s3}$ &394&1357\\
$\Gamma_{s4}$& 0.16& 0.80&$A_{2}$  &3500&6789\\
$\Gamma_{s5}$& 0.54& 0.80&$A_{s5}$ &361&1351\\
$\Gamma_{sss1}$& 0.24& 1.17&$A_{sss1}$ &-938&1477\\
$\Gamma_{sss2}$& 0.26& 1.17&$A_{sss2}$ &-763&1671\\
$m_{Q_c}^2$ & $(1890)^2$ & $(2252)^2$ & 
$m_{Q_b}^2$ & $(3110)^2$ & $(1934)^2$   \\ 
$m_{Q_a}^2$ & $(3110)^2$ & $(1934)^2$ & 
$m_{u_c}^2$ & $(2010)^2$ & $(2352)^2$\\ 
$m_{u_b^{'}}^2$ & $(1460)^2$ & $(4382)^2$ & 
$m_{d_b}^2$ & $(1500)^2$ & $(2172)^2$\\
$m_{d_d}^2$ & $(4370)^2$ & $(6155)^2$ & 
$m_{d_c}^2$ & $(1460)^2$ & $(4382)^2$\\ 
$m_{\bar{h}_c}^2$ & $(531)^2$  & $(4624)^2$ &
$m_{\bar{h}_d}^2$ & $(1040)^2$ & $(4549)^2$\\
$m_{\bar{h}_a}^2$ & $(1250)^2$ & $(1817)^2$ &
$m_{h_a}^2$ & $(1390)^2$ & $(1828)^2$\\
$m_{h_c}^2$ & $-(311)^2$ & $(4177)^2$ &
$m_{h_d}^2$ & $(1400)^2$ & $(1833)^2$\\
$m_{h_g}^2$ & $(4950)^2$ & $(7639)^2$ &
$m_{h_b}^2$ & $-(299)^2$ & $(4939)^2$\\ 
$m_{e_a}^2$ & $(1920)^2$ & $(2590)^2$ & 
$m_{e_f}^2$ & $(1890)^2$ & $(2590)^2$\\ 
$m_{e_b}^2$ & $(1750)^2$ & $(1764)^2$ & 
$m_{e_e^{'}}^2$ & $(1500)^2$ & $(5314)^2$\\ 
$m_{e_i}^2$ & $(1750)^2$ & $(1764)^2$ & 
$m_{\varphi_{20}}^2$ & $-(715)^2$ & $(5177)^2$\\ 
$m_{\varphi_{21}}^2$ & $-(718)^2$ & $(6328)^2$ & 
$m_{\varphi_{25}}^2$ & $(1630)^2$ & $(4337)^2$\\ 
$m_{\varphi_{28}}^2$ & $(4470)^2$ & $(5982)^2$ &
$m_{\varphi_{29}}^2$ & $(2210)^2$ & $(6398)^2$\\
\hline\hline
\end{tabular}
\end{center}
\noindent  Table VI: $P_2P_3|_{F}$ flat direction: values of the
parameters (for the observable sector) at $M_{String}$ and $M_{Z}$, with
$M_{Z'}=1008$ GeV and $\alpha_{Z-Z'}=0.004$. All mass parameters are
given in GeV.

\end{document}